\documentclass[twocolumn,showpacs,preprintnumbers,amsmath,amssymb]{revtex4}

\usepackage{amssymb}
\usepackage{mathrsfs}

\usepackage{graphicx}
\usepackage{dcolumn}
\usepackage{bm}
\usepackage{mathrsfs}

\begin{document}

\title{A Markov chain Monte Carlo analysis to constrain dark matter properties with directional detection}

\author{J. Billard}
\email{billard@lpsc.in2p3.fr}
  
\author{F. Mayet}

\author{D. Santos}
\affiliation{Laboratoire de Physique Subatomique et de Cosmologie, Universit\'e Joseph Fourier Grenoble 1,
  CNRS/IN2P3, Institut Polytechnique de Grenoble, Grenoble, France}

%
%
\date{\today}

\begin{abstract}
Directional detection is a promising dark matter search strategy. Indeed, weakly interacting massive particle (WIMP)-induced recoils would 
present a direction dependence toward the Cygnus constellation, while 
background-induced recoils exhibit an   isotropic distribution in the Galactic rest frame. Taking advantage of these characteristic features and even in the presence of a sizeable background, 
it has recently been shown that data from forthcoming directional detectors  could lead  either to a competitive exclusion or to a conclusive discovery, depending on  
the value of the   WIMP-nucleon cross section.  However, it is possible to further exploit these upcoming data by using the strong dependence of the WIMP signal 
with : the WIMP mass and the local WIMP velocity distribution. Using a Markov chain Monte Carlo analysis of recoil events, we show  for the first time the possibility to 
constrain the unknown WIMP parameters, both from particle physics (mass and cross section) and Galactic halo (velocity dispersion along the three axis), 
leading to an identification of non-baryonic dark matter.
\end{abstract}

\maketitle

%

\section{Introduction}
Directional detection of Galactic dark matter  has been first proposed by 
D.~N.~Spergel \cite{spergel} highlighting the fact that even low angular resolution directional detectors 
could be used to show a clear asymmetry in the forward/backward distribution of weakly interacting massive particle (WIMP) events with respect 
to the direction of the Cygnus constellation.\\
Beyond the simple asymmetry feature, it has recently been shown that dedicated statistical data analysis of forthcoming directional detectors~\cite{white,mimac,drift,dmtpc,newage} could lead 
 either to a competitive exclusion \cite{billard.exclusion} or to a conclusive discovery \cite{billard.disco,green.disco}, depending on the value of the  
 WIMP-nucleon cross section. In the latter case, by using a map based likelihood analysis and even in the presence of a sizeable background,  it is possible to 
 show that the main incoming direction does correspond to the direction of the Cygnus constellation 
 ($\ell_{\odot}, b_{\odot}$).
   This is indeed the discovery proof of this detection
strategy and it has been shown that a 10 kg $\rm CF_4$ detector (MIMAC) operated during 3 years, would  allow for  a high significance discovery down to $\sigma^{SD} \simeq 10^{-4} \ {\rm pb}$ 
\cite{billard.idm2010}. In this paper, we strive to go one step beyond by 
trying to constrain the properties of Galactic dark matter with directional detection.\\

Indeed, constraining WIMP parameters (mass $m_\chi$ and cross section $\sigma_n$) with upcoming dark matter experiments is a main concern of 
current phenomenological studies,  using either  indirect detection \cite{bernal,bernal2}, direct detection on its own 
\cite{bernal2,green.masse1,green.masse2,Drees:2007hr,drees.masse,Shan:2010hr}, in combination with collider data~\cite{Bertone:2010rv} or  with 
  the measurements of halo star  kinematics \cite{Strigari:2009zb}. The quest for a model-independent formalism is a difficult task 
  as the signal expected in direct detection depends  on  the properties of both 
  the WIMP particle (mass and cross section) and  the Galactic dark matter halo 
  (three-dimensional local WIMP velocity distribution and density). This approach is of particular interest in the context of
  competitive upcoming experiments which might be able to give positive WIMP detection instead of background rejection. 
  M.~Drees and C.~L.~Shan have proposed a model-independent reconstruction of the WIMP velocity distribution as well as its various moments
  (mean velocity, dispersions, ...), providing the WIMP mass is {\it a priori} known  \cite{Drees:2007hr} or deduced from positive signals from
  at least two direct detectors with different target nuclei \cite{drees.masse}. 
  The complementary approach is to constrain the WIMP properties with the help of a high dimensional multivariate analysis and within
   the framework of a general halo model, with a large number of parameters. Thus, the main strength of this study, and hence of directional detection, is the possibility of 
   constraining the properties of both the dark matter particle and the dark matter halo with a single experiment. The choice of the fitting model 
   must be well motivated {\it e.g.} by N-body simulations, as it remains as an {\it ansatz}.\\

Directional detection presents a high identification potential thanks to the use of the double-differential spectrum 
${\mathrm{d}^2R}/{\mathrm{d}E_R\mathrm{d}\Omega_R}$, also called the directional event rate, in a given recoil energy range. 
Indeed, its shape    depends both on the WIMP mass and WIMP velocity distribution, while 
the magnitude mainly depends on  the product of the  local WIMP density and the WIMP-nucleon cross
section.  Within the framework of a multivariate recoil event analysis using 
a Markov chain Monte Carlo (MCMC), we show for the first time the possibility to 
constrain, with a single directional experiment, the unknown WIMP parameters, both from particle physics ($m_\chi, \sigma$) 
and Galactic halo (velocity dispersion along the three axis), leading to an identification of non-baryonic dark matter. It is, of course, possible to include external data,
{\it e.g.} halo star kinematics as in \cite{Strigari:2009zb}, and to relax some astrophysical inputs, as $\rho_0$ for instance. However, in this work, we focus on the
contribution of directional detection on its own, highlighting the need for future large directional detectors.\\

The paper is organized as follows. In Sec. II, the  dark matter halo modeling is introduced while 
 the directional detection framework is presented in Sec. III. Then, the Markov chain Monte Carlo analysis is detailed in Sec. IV, highlighting the performance of such a method in the context of
 directional detection. Sec. V presents the results of this 8 parameter analysis for a directional detector with a sizeable background contamination and in the case of a benchmark dark matter model.
 Departures from this input model, by changing the WIMP mass, the velocity anisotropy and the background assumptions are presented in Sec. VI.

\section{dark matter halo modeling}
\label{halomodeling}
Direct detection depends   crucially on the local WIMP velocity distribution 
\cite{alenazi.directionnelle,green.halo,Serpico:2010ae}  and it is important to investigate the effect of 
halo modeling on exclusion limits and allowed regions. The alternative strategy is to build 
a multivariate analysis, using a halo model with a large number of parameters to be constrained by the analysis. 
The isothermal sphere halo model is often considered but it is worth going beyond the standard assumption
 especially when considering recent hints in favor of triaxiality.\\
Indeed, recent  results from 
N-body simulations are in favor of triaxial dark matter halos with anisotropic velocity distributions 
and potentially containing substructures as subhalos (clumps) and dark disk \cite{Ling:2009cn,Nezri,Bruch:2008rx,Read:2008fh}. 
Moreover,  recent observations of Sagittarius stellar
tidal stream have shown evidence for a triaxial Milky Way dark matter halo \cite{Law:2009yq}, with the 
short axis being approximately aligned with the Galactic $\hat{x}$ axis (toward the Galactic center), and the longest 
with the Galactic $\hat{y}$ axis (in the direction of the solar motion).  However, 
it is noteworthy that this result holds true at large radius (60 kpc) and N-body simulations have shown that there can be significant
variations of the axis ratios with radius \cite{Hayashi}. Hitherto, there is no 
observational evidence of triaxiality at solar position.\\

The multivariate Gaussian WIMP velocity distribution has been first 
proposed by N. W. Evans {\it et al.} \cite{Evans:2000gr}. It corresponds  to the simplest triaxial generalization of the standard isothermal sphere with a density profile 
$\rho(r)\propto 1/r^2$, leading to a smooth WIMP velocity distribution without substructure, with a flat rotation curve and in dynamical equilibrium. 
The velocity dispersion tensor $\boldsymbol\sigma_v$, given by the Jeans equations, is symmetric. Thus, one can find an orthogonal basis in which the tensor is diagonal
 leading to the following expression of
the WIMP velocity distribution in  the solar system rest frame,
\begin{equation}
f(\vec{v}) = \frac{1}{(8\pi^3\det{\boldsymbol\sigma}^2_v)^{1/2}}\exp{\left[-\frac{1}{2}(\vec{v} - \vec{v}_{\odot})^T {\boldsymbol\sigma}^{-2}_v(\vec{v} - \vec{v}_{\odot})\right]}
\label{WIMPVelocity}
\end{equation}
where the velocity dispersion tensor ${\boldsymbol\sigma}_v = \text{diag}[\sigma_{x}, \sigma_{y}, \sigma_{z}]$ is 
assumed to be diagonal in the Galactic rest frame  and $\vec{v}_{\odot}$ is the Sun motion with respect to
the Galactic rest frame.\\
The velocity anisotropy $\beta(r)$, is then defined as \cite{biney},
  \begin{equation}
  \beta(r) = 1 - \frac{\sigma^2_{y} + \sigma^2_{z}}{2\sigma^2_x}
  \label{eq:beta}
  \end{equation}
  According to N-Body simulations with or without
  baryons  \cite{Nezri,vogelsberger,kuhlen,Moore:2001vq}, the $\beta$ parameter at $R_{\odot} = 8$ kpc of the Galactic center, spans the 
  range $0-0.4$ which is in favor of radial anisotropy.

As stated above, our choice is to develop a high dimensional multivariate analysis considering a general enough halo model, {\it i.e.} with a large number of parameters,
 and  by constraining
all of them  with the data analysis of a single experiment. The choice of the halo model must be 
carefully done as it remains as an ansatz. Following recent results from N-body simulations with baryons \cite{Nezri,Teyssier:2001cp}, the choice  
 of  a multivariate Gaussian seems to be a reasonable guess, although one could argue that deviations are observed in the WIMP 
velocity distribution,  making it closer to a generalized Gaussian  or
 even a double Maxwellian distribution when considering the  presence of a corotating dark disk. We argue that 
 worrying about the
  exact shape of the WIMP velocity distribution seems to be not relevant, in particular when taking 
  into account the fact that the resolution of current numerical simulations is many
  orders of magnitude larger than the scale of the ultralocal dark matter distribution 
  probed by current and future detectors.   
 This is why we have chosen a multivariate Gaussian WIMP velocity distribution as a fitting model, in  
  a first attempt to constrain both the WIMP parameters ($m_{\chi},\sigma_n$) and the dark matter halo properties
 using directional detection. Effect of nonsmooth halo model with substructures and/or streams will be addressed in a forthcoming paper.  \\

In the following, 
the input halo model used to generate simulated data, is chosen according to two models : a standard isotropic halo ($\beta=0$) in which case the velocity dispersions are
linked to the local circular velocity $v_0 = 220$ km/s as $\sigma_{x} = \sigma_{y} = \sigma_{z} = v_0/\sqrt{2} \approx 155$  km/s;
and an  anisotropic halo ($\beta = 0.4$), 
with the following velocity dispersions $\{\sigma_x = 200$ km/s; $\sigma_z = 169$ km/s; $\sigma_y = 140$ km/s$\}$. 
The latter case corresponds to  the logarithmic ellipsoidal halo model from \cite{Evans:2000gr} with the 
Sun located on the major axis of the halo, the axis ratios $p$ and $q$ being equal to 0.9
and 0.8, respectively. This is usually taken as an extreme case 
for the anisotropy, in order to avoid instabilities arising  
when the ratio of any of the velocity dispersion is greater than 3.  Indeed, as
discussed in \cite{green.halo,Evans:2000gr}, 
in order to consider only physically relevant model,  every velocity dispersions has to 
satisfy the following constraint: $\sigma_{j}/3 < \sigma_{i} < 3\sigma_{j}$.


\section{Directional detection framework}
 
\subsection{Detector configuration}
\label{sec:dec}
Several dark matter directional detectors \cite{white} are being developed and/or operated : 
MIMAC~\cite{mimac}, DRIFT~\cite{drift}, DM-TPC~\cite{dmtpc} and NEWAGE~\cite{newage}. Directional detection   requires 3D track reconstruction  of
 recoiling nuclei down to a few keV with sense recognition. {\it In fine}, an ideal directional detector should allow one to evaluate 
the double-differential spectrum $\mathrm{d}^2R/\mathrm{d}E_R\mathrm{d}\Omega_R$ in a given recoil energy range  [$E_{R_1},E_{R_2}$]. The lower bound is due to the threshold 
ionization energy taking into account the
 quenching factor, while the upper bound allows one to limit  background contamination, as most of the WIMP events are concentrated at low-recoil energy.\\
 In the following, we consider an ideal detector configuration  which could be within reach in a few years. The
 configuration of the MIMAC project is chosen :  
 a 10 kg $\rm CF_4$ detector, operated at 50 mbar and 
 allowing 3D reconstruction of recoiling tracks with sense recognition. The chosen recoil energy range is between 5 and 50 keV 
 and an exposure $\xi = 30 \ {\rm kg.year}$ is taken into account for data simulation. In order to treat realistic cases, 
 we allow for a sizeable  residual background contamination in the data. Indeed, the 
 discrimination of isotropic background events from WIMP events has been early recognized as the main strength of this detection strategy
 \cite{spergel}. However, as discussed in 
 \cite{billard.disco,billard.exclusion,bernal2,green.masse1,green.masse2,drees.masse,Shan:2010hr}, one of the key issues for direct detection 
 is the unknown background energy distribution. 
Two extreme cases may be considered  \cite{bernal2,green.masse1,green.masse2,drees.masse,Shan:2010hr}  : flat or exponentially decreasing with 
increasing recoil energy, {\it i.e.} with the same feature as the WIMP-induced energy spectrum. Within the framework of a dedicated statistical data analysis
 aiming at the  
 identification of dark matter, residual background should be accounted for and we will show that it does only mildly alter the result.\\
Energy and angular resolutions are other points to be carefully handled. However, it depends on various track parameters such as track length, gaz mixture,   
initial track position and direction. A full study of 3D track reconstruction is underway \cite{billard.track} and we argue that taking into account
finite angular resolution required a full coupling of track reconstruction analysis with this MCMC method. As a first step, and as our goal is to show the identification
potential of directional detection, an ideal detector is considered hereafter, {\it i.e.} perfect energy and angular resolutions.
As shown in \cite{billard.exclusion}, the effect of finite angular/energy resolution has been shown to be small as far as  directional exclusion limits are concerned, 
providing the angular resolution is well estimated via  detector commissioning, e.g. by using a neutron field \cite{golabek}.

\subsection{Directional detection}
\label{sec:directional}
The detector velocity in
the Galactic rest frame  corresponds to  $\vec{v}_{\odot}$, when neglecting the Sun peculiar velocity and the Earth orbital 
velocity about the Sun\footnote{Obviously, for real data analysis, these two
components of the detector velocity have to be considered in order to have an accurate analysis.}.  
We consider the value $\vec{v}_{\odot} = 220$ km.s$^{-1}$ along the $\hat{y}$ axis. In such case, 
the main incoming direction of the WIMP signal should
be pointing toward ($\ell_{\odot} = 90^{\circ},b_{\odot} = 0^{\circ}$). 
Using the Galactic coordinates ($\ell$, $b$), the WIMP velocity is written in the Galactic rest frame as:
\begin{equation}
\vec{v} = v(\cos\ell\cos b \ \hat{x} + \sin\ell\cos b \ \hat{y} + \sin b \ \hat{z})
\end{equation}
Following \cite{gondolo}, the
directional recoil rate is given by,
\begin{equation}
\frac{\mathrm{d}^2R}{\mathrm{d}E_R\mathrm{d}\Omega_R} = \frac{\rho_0\sigma_0}{4\pi m_{\chi}m^2_r}F^2(E_R)\hat{f}(v_{\text{min}},\hat{q}),
\label{directionalrate}
\end{equation}
with $m_{\chi}$ the WIMP mass, $m_r$ the WIMP-nucleus reduced mass, $\rho_0 = 0.3$ GeV/c$^2$/cm$^3$  the local dark matter density (see Sec.~\ref{sec:WIMPParam}
 for discussion), $\sigma_0$   the WIMP-nucleus elastic scattering cross section, $F(E_R)$  the form factor, $\hat{q}$ refers to the recoil direction expressed in the
 Galactic coordinates  and  
$v_{\text{min}} = \sqrt{m_NE_R/2m^2_r}$ is the   minimal WIMP velocity required to produce a
nuclear recoil of energy $E_R$.  In the case of an axial coupling and within the Born approximation, 
the expression of the form factor is given by  \cite{lewin}~:
\begin{equation}
F(E_R) = \frac{\sin \Big[\sqrt{2m_NE_R}\times R(^{A}X)\Big]}{\sqrt{2m_NE_R}\times R(^{A}X)}
\end{equation}
where $R(^{A}X)$ is the radius of the target nucleus.\\ 
Eventually, $\hat{f}(v_{\text{min}},\hat{q})$ is the three-dimensional Radon transform \cite{radon} of the WIMP 
velocity distribution $f(\vec{v})$ defined as,
\begin{equation}
\hat{f}(v_{\text{min}},\hat{q}) = \int d^3v \ \delta(v_{\text{min}} - \vec{v}.\hat{q})f(\vec{v})
\end{equation}
Geometrically, the Radon transform is the integral of the function $f(\vec{v})$ on a plane orthogonal to the direction $\hat{q}$ 
at a distance $v_{\text{min}}$ from the
origin. Using the Fourier slice theorem, P.~Gondolo found the expression of the 
Radon transform of the multivariate Gaussian to be \cite{gondolo},
\begin{equation}
\hat{f}(v_{\text{min}},\hat{q}) = \frac{1}{(2\pi\hat{q}^T{\boldsymbol\sigma}^2_v\hat{q})^{1/2}}\exp{\left[-\frac{\left[v_{\text{min}} - \hat{q}.\vec{v}_{\odot}\right]^2}{2\hat{q}^T{\boldsymbol\sigma}^2_v\hat{q}}\right]}.
\end{equation}
Together with (\ref{directionalrate}), this expression is of particular interest in the context of massive MCMC 
calculations, as it allows one to avoid time-consuming  evaluation 
of the 3D integral  of $f(\vec{v})$ for each event at each step. It is, however, equivalent to 
the directional recoil rate of \cite{copi}.\\

In the energy range $[E_{R_1}, E_{R_2}]$, the expected number of WIMP events $\mu_s$ corresponding to a given set of 
physical parameters is given by,
\begin{equation}
\mu_s = \xi\int^{E_{R_2}}_{E_{R_1}}\int_{\Omega_R} \ \frac{\mathrm{d}^2R}{\mathrm{d}E_R\mathrm{d}\Omega_R} \ \mathrm{d}E_R\mathrm{d}\Omega_R
\end{equation}
where $\xi$ is the total exposure.

\section{The directional Markov chain Monte Carlo method}
\label{sec:mcmc}
We present a new method, based on a
 Markov chain Monte Carlo analysis, to extract information on dark matter from directional data. 
First, the interest of use of MCMC algorithm is outlined.  Then, the method is fully described in the following section. 
Discussion on chain efficiency will be done in order to prove that MCMC algorithms are suited for this type of analysis.

\subsection{Interest of the MCMC algorithm}
As stated above,  directional detection offers the possibility of using three-dimensional datasets: the recoil energy $\rm E_R$ and its direction
in Galactic coordinates ($\rm \ell_R, b_R$). It follows that  constraints on dark matter properties should be enhanced by the use of 
directional detection when compared to directional insensitive detection. This is of particular interest when developing a high dimensional multivariate analysis 
aiming at going beyond the standard isotropic
halo assumption. Indeed, the information enclosed in the directional event rate ($d^2R/dE_Rd\Omega_R$), {\it i.e.} the energy spectrum and the 2D shape of the 
angular distribution, allows one in principle to increase the number of degrees of freedom of the fitting model. 
Indeed, adding directional information to the energy one  allows one to remove degeneracies among  
fitting parameters and hence to deduce consistent constraints.\\
In the following, we list the free parameters of our fitting model  :
\begin{itemize}
\item ($\sigma_{x}$, $\sigma_{y}$, $\sigma_{z}$)  the three velocity dispersions  of the   
local WIMP velocity distribution,    
\item   ($\ell_{\odot}$, $b_{\odot}$) referring to 
the main direction of the recoiling nuclei. It is indeed an unambiguous signature of dark matter detection \cite{billard.disco}, 
\item  $m_{\chi}$ the WIMP mass, 
\item  $\sigma_n$ the WIMP-nucleon cross section directly related to $\sigma_0$ in the 
pure proton approximation for the fluorine target,  
\item  $R_b$ the  background event rate in the considered energy range [5, 50] keV. The background events are those remaining after the electron/nuclear recoil rejection,
based for instance on the length/energy  discrimination \cite{Grignon}.
\end{itemize}
This leads to an eight-parameter analysis of directional dark matter dataset allowing us to quantitatively constrain the 
WIMP properties and the dark matter halo profile. Prior ranges are presented in Table~\ref{tab:prior}.\\

\setlength{\tabcolsep}{0.1cm}
\renewcommand{\arraystretch}{1.4}
\begin{table}[b]
\begin{center}
\hspace*{-0.5cm}
\begin{tabular}{|c|c|}
\hline
 Parameter  &   Prior range \\ \hline \hline  
 $m_{\chi} \ {\rm (GeV/c^2)}$  &  $(5,1000)$ \\ \hline  
 $\log_{10}(\sigma_n \ {\rm (pb))}$ & $(-5,-1)$ \\ \hline 
 $\ell_{\odot} \ {\rm (^\circ)}$ & $(-180,+180)$ \\ \hline 
 $b_{\odot} \ {\rm (^\circ)}$ & $(-90,+90)$ \\ \hline 
 $\sigma_{x,y,z} \ {\rm (km.s^{-1})}$ & $(5,500)$ \\ \hline  
 $R_b \ {\rm (kg^{-1}year^{-1})}$ & $(0,50)$ \\ \hline  
\end{tabular}
\caption{Parameters with their uniform prior ranges used for all MCMC analysis.}
\label{tab:prior}
\end{center}
\end{table}
\renewcommand{\arraystretch}{1.1}

As the number of free parameters is large, grid calculation of likelihood or $\chi^2$ functions are not suitable due to the exponential growth of the volume of the parameter space. Indeed,
 in order to ensure a scan of all the physical parameter space, the {\it regions of interest}, {\it i.e.} the region where 
 the model fits the data, will fill only a tiny part of the whole 
 volume. This corresponds to a waste of computation time that is avoided using   MCMC algorithm. Indeed, Markov chains are 
 used in order to sample the likelihood (or $\chi^2$)
 distribution according to Bayesian statistics, enabling the enlargement of the parameter space 
 at a minimal computing time cost by focusing on the regions of interest.\\
 In the following we provide a brief description of the MCMC, emphasizing its  use in the context of directional detection. We refer the
 reader to a more complete description, within the framework of cosmic ray physics, in \cite{putze} and references therein.

\subsection{Description of the method}

In a general description, an $m$-dimensional parameter space is described by the following basis $\vec{\theta} = \{\theta^{(1)}, \theta^{(2)},...,\theta^{(m)}  \}$, 
where each element $\theta^{(\alpha)}$ refers to one of the  physical parameter of interest. 
The MCMC algorithm   enables us to sample the conditional posterior {\it Probability Density Functions} (PDF) of each parameter given the data
 $P(\vec{\theta}|\vec{D})$, where $\vec{D}$ refers to the number of events $N$, their direction ($\ell_R, b_R$) and their energy $E_R$. This can be achieved with
 the use of the Bayes' theorem applied to parameter inference,
 \begin{equation}
 P(\vec{\theta}|\vec{D}) = \frac{P(\vec{D}|\vec{\theta}) \times \Pi(\vec{\theta})}{P(\vec{D})}
 \end{equation}
where $P(\vec{D})$ is the data probability, also called the {\it evidence}, which can be regarded as a normalization factor, $\Pi(\vec{\theta})$ is the prior probability
indicating the degree of belief   before observing the data. Finally, $P(\vec{D}|\vec{\theta})$ corresponds to the likelihood function written 
$\mathscr{L}(\vec{\theta})$. In this framework, the posterior PDF $P(\vec{\theta}|\vec{D})$ is the normalized product of the likelihood function with the priors. Using a Bayesian
approach, the posterior PDF of each single parameter $\theta^{(\alpha)}$ is given by the marginalisation of the multidimensional $P(\vec{\theta}|\vec{D})$ distribution
 over the other parameters $\theta^{(\beta\neq\alpha)}$,
\begin{equation}
P(\theta^{(\alpha)}|\vec{D}) = \int_{\Omega_{\beta}, \ \forall \beta \in [1,m]\backslash \{\alpha\}} P(\vec{\theta}|\vec{D}) \ \mathrm{d}\theta^{(\beta)}.
\end{equation}
From each one-dimensional PDF, we can estimate the expected value of a given parameter and its confidence level (CL).
 The difficulty is then 
to evaluate the multidimensional target PDF $ p(\vec{\theta}) \equiv P(\vec{\theta}|\vec{D})$. For the above-mentioned reasons, instead of using a grid calculation algorithm, 
we developed a MCMC algorithm in order to evaluate $ p(\vec{\theta})$.
This Monte Carlo sampling of the target function is done using Markov chains which are a sequence of $N$ points in the $m$-dimensional parameter space,
\begin{equation}
\{ \vec{\theta}_i \}_{i=1,...,N} \equiv \{\vec{\theta}_1,\vec{\theta}_2,...,\vec{\theta}_N\}
\end{equation}
 which is constructed according to the Metropolis-Hastings algorithm ensuring that the stationary distribution of each chain corresponds to the target distribution
 $p(\vec{\theta})$ being sampled. The Metropolis-Hastings algorithm is a random walk in the parameter space where each step $\vec{\theta}_{i+1}$ is derived from the step $\vec{\theta}_{i}$
 with the following procedure :

 \begin{itemize}
 \item At each step $\vec{\theta}_{i}$ a trial step $\vec{\theta}_{\text{trial}}$ is generated from a proposal distribution $q(\vec{\theta}_{\text{trial}}|\vec{\theta}_{i})$.
 \item This trial step is accepted or not according to the acceptance probability $a$ calculated as follows :
 \begin{equation}
 a = a(\vec{\theta}_{\text{trial}}|\vec{\theta}_{i}) = \text{min}\left(1,\frac{p(\vec{\theta}_{\text{trial}})}{p(\vec{\theta}_{i})}\frac{q(\vec{\theta}_{\text{trial}}|\vec{\theta}_{i})}{q(\vec{\theta}_{i}|\vec{\theta}_{\text{trial}})}  \right)
 \label{acceptance}
 \end{equation}
 The probability for the trial step to be  accepted is equal to $a$. Notice 
 that in the case of a symmetric  proposal function $q$ we have  
 $q(\vec{\theta}_{\text{trial}}|\vec{\theta}_{i}) = q(\vec{\theta}_{i}|\vec{\theta}_{\text{trial}})$, which simplifies the expression of eq.(\ref{acceptance}).

 \item If the trial step is accepted, then  $\vec{\theta}_{i+1} = \vec{\theta}_{\text{trial}}$ and if not, the chain stagnate at the same point in the parameter space leading to
 $\vec{\theta}_{i+1} = \vec{\theta}_{i}$.
 \end{itemize}

 \begin{figure}[t]
\begin{center}
\includegraphics[scale=0.4,angle=0]{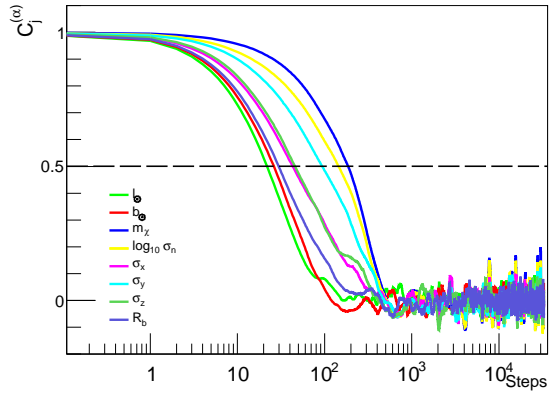}
\caption{Correlation function of the 8 parameters from the MCMC run applied to our benchmark model (see sec.~\ref{sec:res.iso}) with the simple multivariate Gaussian
proposal function.} 
\label{fig:CorrelationLength}
\end{center}
\end{figure}

Three characteristics of the Markov chains are worth being investigated in order to ensure a consistent sampling of the target function :\\

{\it Burn-in length (b):} it corresponds to the number of   steps (or iterations) to be 
removed from the beginning in order to {\it forget} the starting point of the random walk.
 It is estimated as the first step reaching the median value of the target distribution $E[p(\vec{\theta})]$ as
\begin{equation}
p(\vec{\theta}_b) > E[p(\vec{\theta})]
\end{equation}

{\it Correlation length (l):} it is the required minimal length between two steps so that they can be considered as uncorrelated. 
By construction, each step depends on the 
previous one. Then, in order to get
independent steps, some {\it subsampling} is needed. It corresponds to rejecting all steps which are closer than $l$ to each other. 
The correlation length $l^{(\alpha)}$
of each parameter $\theta^{(\alpha)}$ is estimated by computing the autocorrelation function $c^{(\alpha)}_j$ where $j$ corresponds to 
the distance between two steps. 
Indeed, $l^{(\alpha)}$ is defined as the smallest $j$ for which the correlation function is strictly less 
than 1/2, {\it i.e} $ c^{(\alpha)}_j < 1/2$. It should be noticed that the limit of 1/2 is arbitrary but has been shown to be sufficient in order to consider the steps
$\theta^{(\alpha)}_i$ and $\theta^{(\alpha)}_{i+j}$ as uncorrelated \cite{putze}.
Then, the correlation length of the whole chain $l$ is defined as,
\begin{equation}
l =  \text{max}[l^{(1)},...,l^{(\alpha)},...,l^{(m)}]
\end{equation}

 \begin{figure*}[t]
\begin{center}
\includegraphics[scale=0.40,angle=0]{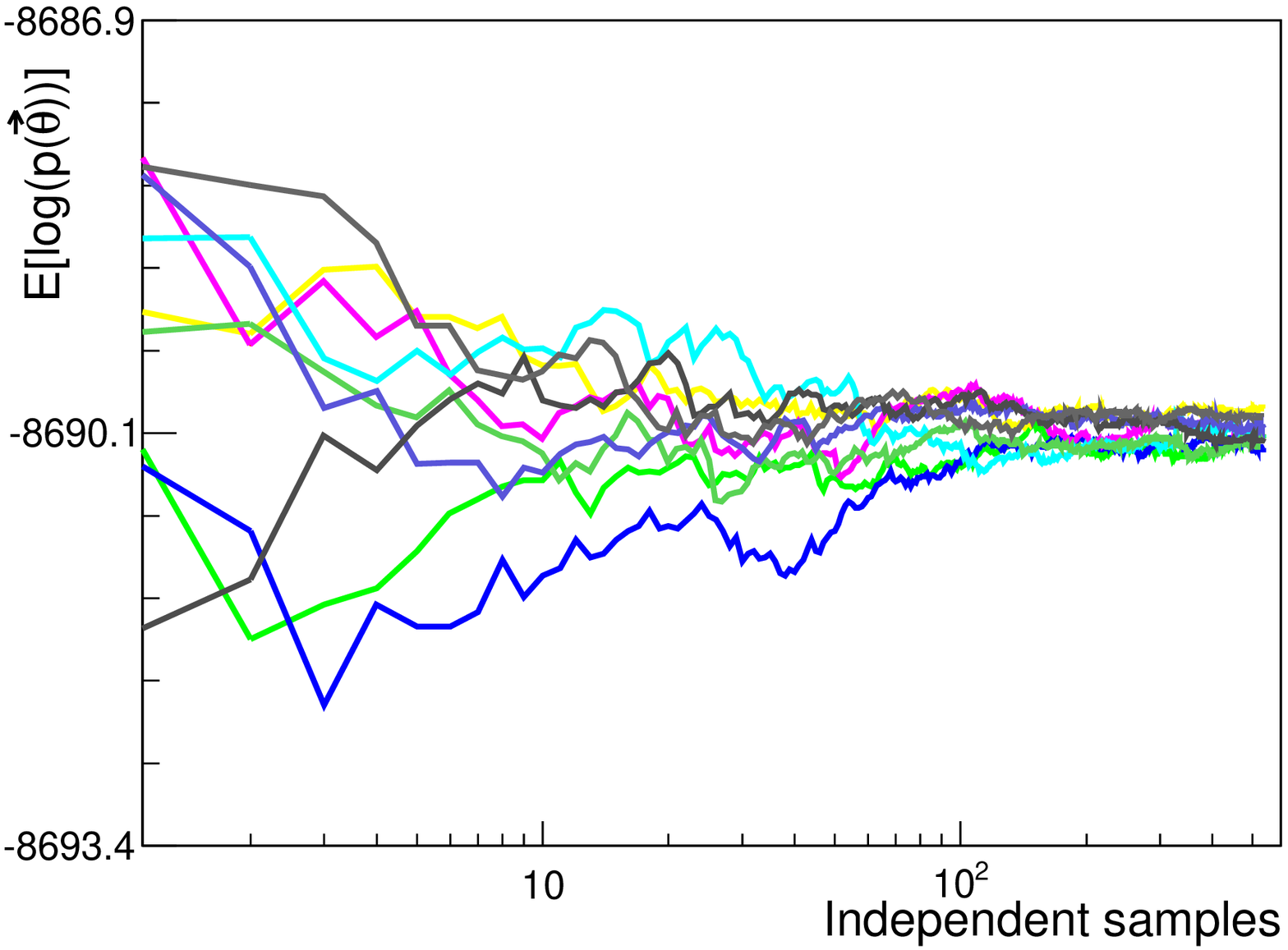}
\hspace{5mm}
\includegraphics[scale=0.40,angle=0]{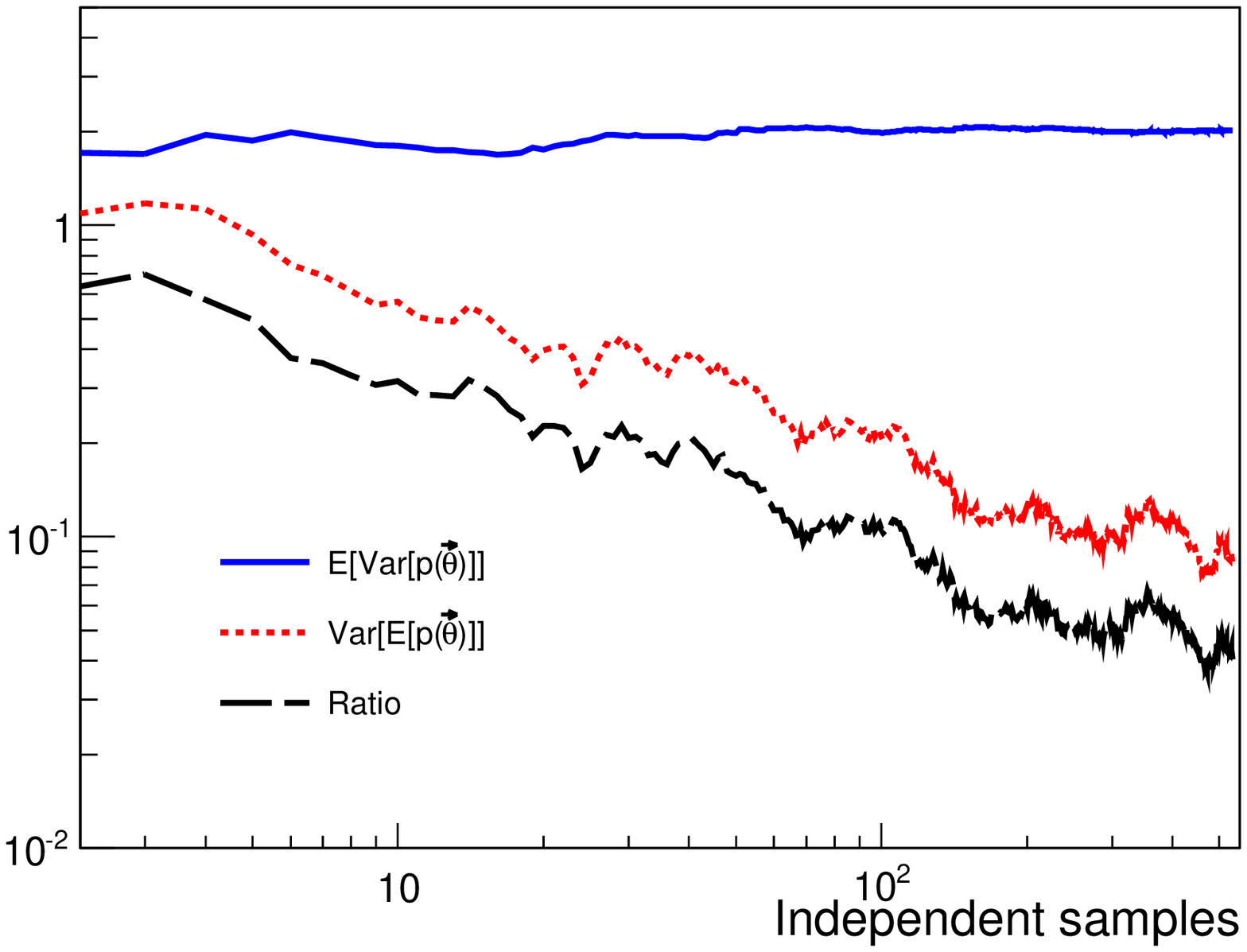}
\caption{Left panel : mean of the log-likelihood value of each Markov chain $\text{E}[\log(p(\vec{\theta}))]$ 
as a function of the number of independent samples  for  10 Markov chains. 
Right panel : $\text{Var}[\text{E}(p(\vec{\theta}))]$, $\text{E}[\text{Var}(p(\vec{\theta}))]$ and convergence ratio $r$ as a function of
the number of steps. Both figures come from the MCMC run applied to our benchmark model (see Sec.~\ref{sec:res.iso}).} 
\label{fig:Convergence}
\end{center}
\end{figure*}

Hence, in order to consider only independent samples (steps) $\vec{\theta}_{\text{ind}}$ we have subsampled each Markov chain according to 
the following procedure  
$\vec{\theta}_{\text{ind}} = \vec{\theta}_{i=b+kl}$ with $k$ being an integer.
Figure \ref{fig:CorrelationLength} represents the autocorrelation function for the eight parameters from a MCMC analysis discussed in Sec.~\ref{sec:res.iso}.
 For this chain, we can see
that the correlation length is equal to 187 due to the WIMP mass parameter. Indeed, as  explained in the following section, the strong correlation between the WIMP
mass and cross-section will induce larger correlation length. Hence, as the correlation length is linked to the {\it stagnation} of the chain, in order to have a
smaller value of $l$, the proposal function has to be carefully chosen to approximate the target PDF.  \\
The efficiency of a Markov chain Monte Carlo sampling can
then be estimated as the fraction of  independent samples $N_{\text{ind}}$ with respect to 
the total number of samples $N$, where $N_{\text{ind}}$ is given by :
 \begin{equation}
 N_{\text{ind}} =  \frac{N-b}{l}
\end{equation}
More efficient is a MCMC sampling, since the number of rejected samples is lower, leading
 to a better estimation of the target PDF.
The quality of  the estimation of the target PDF is directly affected by the MCMC sampling efficiency and hence by the 
burn-in and correlation lengths. Depending on the input values of the different parameters
  used to simulate pseudodata of single directional detection experiments, the sampling efficiency is between 0.6\% and 8\%. 
  It mainly
  depends on the correlation lengths which were found to be between 6 and 130 using the second proposal function :
   the multivariate Gaussian with covariance matrix (see below). The sampling efficiency could be enhanced by using other proposal functions not necessarily Gaussian like the
   {\it Binary Space Partitioning} first introduced in MCMC sampling by A.~Putze  {\it et al.} \cite{putze}. However, as we are running a large number of Markov chains in
   parallel with a low computational time, such efficiencies are largely enough to get   well sampled PDFs. \\

{\it chain convergence:} it is a key criteria worth being investigated for MCMC sampling as it ensures that the target 
PDF is being sampled by the different chains. Indeed, the left panel of Fig.~\ref{fig:Convergence} presents the evolution of 
the mean of the 
log-likelihood value  of each Markov chain $\text{E} [ \log(p(\vec{\theta})) ]$ 
as a function of the number of independent samples for  10 Markov chains. From this figure, we can appreciate 
the fact that the mean value of each Markov chain is 
 converging to the same value as well as the variance, not shown here. Then, in order to 
 quantify the convergence, we can form the following ratio :
\begin{equation}
r = \frac{\text{Var}\left[\text{E}(p(\vec{\theta}))\right]}{\text{E}\left[\text{Var}(p(\vec{\theta}))\right]}
\end{equation}
As   seen on the right panel of Fig.~\ref{fig:Convergence}, $\text{Var}[\text{E}(p(\vec{\theta}))]$ tends to 0 whereas 
$\text{E}[\text{Var}(p(\vec{\theta}))]$ tends to a finite number leading to $r \rightarrow 0$ when 
increasing the number of independent samples.
  Then, we can arbitrarily fix a limit $r_c$ which will correspond to a chain convergence if $r<r_c$. Following \cite{MCMCTriaxial}, we have chosen $r_c = 0.2$.
Then, from the Fig.~\ref{fig:Convergence}, we can see that for this MCMC run with 10 independent chains, the convergence status is reached at the 40$^{\text{th}}$
 independent step (sample). However, even if the chain convergence criteria is reached with only a few tenth of independent samples, one should have longer chains of
 independent samples to get a very precise estimation of the target function. In our case we have run between 10 and 100 Markov chains 
 of $10^5$ steps in order to get more
 than $5 \times 10^4$ independent samples for each analysis. Indeed, another interest in computing several Markov chains in parallel is that we can add all the independent
 samples from every chain together to enhance the estimation of the target function. \\

 \begin{figure*}[p]
\begin{center}
\includegraphics[scale=1.2,angle=90]{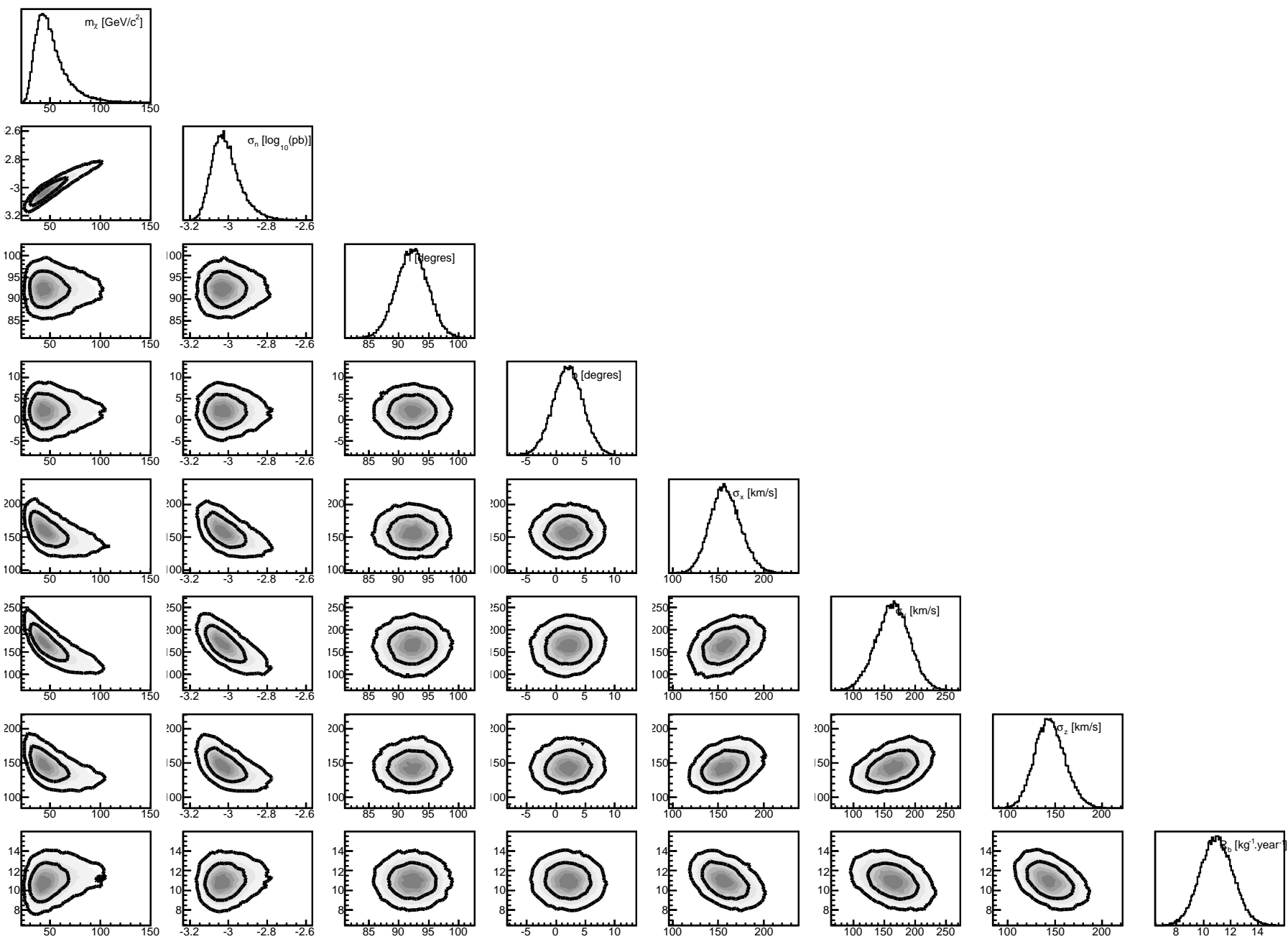}
\caption{
Marginalized distributions (diagonal) and 2D correlations (off-diagonal) plots  of the 8 
parameters from the analysis of simulated data in the case of 
an isothermal halo with a WIMP mass of 50 GeV.$c^{-2}$ and a  WIMP-nucleon cross section 
$\sigma_n = 10^{-3}$ pb.} 
\label{fig:fat50iso}
\end{center}
\end{figure*}

 In this paper, we have considered flat prior for each parameter 
 $\{m_{\chi}, \log_{10}(\sigma_n), \ell_{\odot}, b_{\odot}, \sigma_{x}, \sigma_{y}, \sigma_{z}, R_b \}$.
 In such case, the Bayes'
 theorem is simplified and the target distribution $p(\vec{\theta})$ reduces to the likelihood function $\mathscr{L}(\vec{\theta})$. The latter is given by the extended
 likelihood function (see Ref. \cite{Cowan}) dedicated to unbinned data as,
  \begin{align}
 \mathscr{L}(\vec{\theta}) & = \frac{(\mu_s + \mu_b)^N}{N!}e^{-(\mu_s + \mu_b)}\ \times \nonumber \\
 & \prod_{n = 1}^{N_{\text{event}}} \left[ \frac{\mu_s }{\mu_s + \mu_b} S(\vec{R}_n)  + \frac{\mu_b }{\mu_s + \mu_b}B(\vec{R}_n)\right ]
 \end{align}
 where $\mu_s$ and $\mu_b = R_b\times\xi$ are the expected number of WIMP events and background events respectively. $\vec{R}_n$ refers to the  
 energy and
 direction  of each event while the functions $S$ and $B$ are the directional event rate of the WIMP events and the background events, respectively. 

 As previously  highlighted, in order to optimize the MCMC sampling efficiency, the proposal function must be as 
    close as possible to the target PDF. Two different successive proposal functions are used :
 \begin{itemize}
 \item A multivariate Gaussian in the same basis of the parameter space with dispersions $\sigma^{(\alpha)}$ taken from a fast evaluation of the likelihood function
  on a grid. In such case, we have,
 \begin{equation}
 \theta^{(\alpha)}_{\text{trial}} = \theta^{(\alpha)}_{i} + \sigma^{(\alpha)}x
 \end{equation}
 where $x$ is a random variable distributed  following the normal distribution $\mathscr{N}(0,1)$.
 \item A multivariate Gaussian with the covariance matrix estimated from the previous run. Then, the next step is calculated using:
  \begin{equation}
 \vec{\theta}_{\text{trial}} = \vec{\theta}_{i} + PC\vec{x}
 \end{equation}
 where $C$ is the eigenvalue of the covariance matrix, $P$ is the matrix of the corresponding eigenvectors and $\vec{x}$ is a vector of
 $m$ random variables distributed following $\mathscr{N}(0,1)$.
 \end{itemize}
 In both cases, as the proposal function is a Gaussian, we are in the case where 
 $ q(\vec{\theta}_{\text{trial}}|\vec{\theta}_{i}) = q(\vec{\theta}_{i}|\vec{\theta}_{\text{trial}})$ which simplifies 
 the expression of the acceptance (Eq.~\ref{acceptance}).


\section{Result for a benchmark input dark matter model}
\label{sec:res.iso}

\setlength{\tabcolsep}{0.1cm}
\renewcommand{\arraystretch}{1.4}
\begin{table*}[t]
\begin{center}
\hspace*{-0.5cm}
\begin{tabular}{|c||c|c|c|c|c|c|c|c|c|}
\hline
  & $m_{\chi} \ {\rm (GeV/c^2)}$ &  $\log_{10}(\sigma_n \ {\rm (pb))}$ & $ \ell_{\odot} \ {\rm (^\circ)}$ & $b_{\odot} \ {\rm (^\circ)}$ 
 & $\sigma_{x} \ {\rm (km.s^{-1})}$ & $\sigma_{y} \ {\rm (km.s^{-1})}$ & $\sigma_{z} \ {\rm (km.s^{-1})}$ & $\beta$ & $R_b \ 
 {\rm (kg^{-1}year^{-1})}$ \\ \hline \hline
Input & 50    &   -3  & 90 & 0 & 155 & 155 & 155 & 0 & 10  \\ \hline
Output &  $51.8^{+5.6}_{-19.4}$    & $-3.01^{+0.05}_{-0.08}$  &  $92.2^{+2.5}_{-2.5}$ & $2.0^{+2.5}_{-2.5}$ & $158^{+15}_{-17}$ & $164^{+27}_{-26}$ & $145^{+14}_{-17}$ & $-0.073^{+0.29}_{-0.18}$ & 
$10.97 \pm 1.2 $   \\ \hline   
\end{tabular}
\caption{Comparison of the values of the parameters for the 
input model and as extracted after the MCMC analysis from the marginalized distributions. 
We quote mean value of the PDF distribution and (68 \% CL) error bars.}
\label{tab:modelinput}
\end{center}
\end{table*}
\renewcommand{\arraystretch}{1.1}

For concreteness, we exemplify this directional MCMC method by studying the
case of a given benchmark input model, {\it i.e.} the standard isothermal sphere with an isotropic velocity distribution with $\beta=0$ 
(see Sec.~\ref{halomodeling} for more details).
A sizeable background contamination ($10 \ {\rm kg^{-1}year^{-1}}$) is accounted for, with a flat energy spectrum.  
We consider a  $50 \ {\rm GeV/c^2}$ WIMP with a  WIMP-nucleon axial cross section 
$\sigma_n = 10^{-3}$ pb. The input model is used to generate simulated data in a 10 kg $\rm CF_4$ detector (as proposed by the MIMAC
collaboration) with a three-year exposition time. These data are then analyzed with the directional MCMC method (Sec.~\ref{sec:mcmc}). 
As stated above, the eight parameters $\{m_{\chi}, \log_{10}(\sigma_n), \ell_{\odot}, b_{\odot}, \sigma_{x}, \sigma_{y}, \sigma_{z}, R_b\}$,  are taken as free parameters in the MCMC analysis, with flat priors, 
thus ensuring that the study is model-independent from a point of view of both  particle physics (WIMP properties) and 
Galactic  physics (halo properties). In particular, 
no previous knowledge of the Galactic dark matter halo is needed, the goal being to extract the posterior PDF 
of all parameters  and   check their
consistencies with the input  model. Departure from isotropy as well as modification of the various parameters of 
the input model will be studied in Sec.~\ref{sec:param}.\\

\begin{figure}[h]
\begin{center}
\includegraphics[scale=0.45,angle=0]{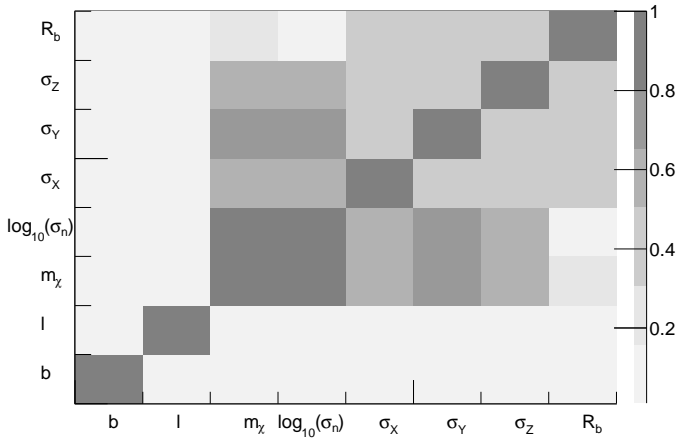}
\caption{Correlation matrix as defined by Eq.~\ref{eq:covmatrix} for the eight parameters of the MCMC analysis in the case of 
an isothermal halo with a WIMP mass of 50 GeV.$c^{-2}$ and a  WIMP-nucleon cross section. The grey scale represents the 
absolute values of $\rho^{\alpha,\beta}$. Signs of correlation can be deduced from Fig.~\ref{fig:fat50iso}.} 
\label{fig:corMatrix}
\end{center}
\end{figure}

Figure \ref{fig:fat50iso} presents marginalized distributions (diagonal) and 2D correlations (off-diagonal) plots of the eight 
parameters of the analysis of simulated data obtained with the benchmark input model. The complete result for each parameter, as extracted from
marginalized distributions, is summarized in  Table~\ref{tab:modelinput}, where the output parameters are characterized by the mean value extracted from their 1D posterior
 PDF, while the error bars are accounted for a 68\% confidence level.
 However, to fully understand correlations between the parameters, the full set of 2D correlations is needed. Moreover, to quantify those correlations among the eight different
 parameters, the correlation matrix defined as
 \begin{equation}
\rho^{\alpha,\beta} = \rho[\theta^{(\alpha)},\theta^{(\beta)}]  =
\frac{\text{cov}[\theta^{(\alpha)},\theta^{(\beta)}]}{\sqrt{\text{var}[\theta^{(\alpha)}]\text{var}[\theta^{(\beta)}]}}
\label{eq:covmatrix}
\end{equation}
  is given in Fig.~\ref{fig:corMatrix} and will be discussed hereafter.\\
  
  The result obtained is threefold   :  the discovery proof is given by the reconstruction of the main incoming direction $(\ell_{\odot},b_{\odot})$ (Sec. \ref{sec:proof}). 
Then, the three velocity dipersions and hence the velocity anisotropy parameter ($\beta$) of the dark matter halo are assessed (Sec.~\ref{sec:velocitydisp})
leading to a constraint
on the properties of the WIMP particle on the ($m_{\chi}, \log_{10}(\sigma_n)$) plane, within the framework of our ansatz.
In the following, we detail the results arising from the analysis of Fig.~\ref{fig:fat50iso}.\\

\subsection{Discovery proof}
\label{sec:proof}
Following a previous study \cite{billard.disco}, we first present the extraction of the main incoming direction of the events ($\ell_{\odot}, b_{\odot}$),
from a pseudodata analysis. This is a blind analysis as these two parameters 
are taken as free parameters of the analysis. It can be concluded 
from marginalized distributions of Fig.~\ref{fig:fat50iso} that the recovered main
recoil direction is pointing towards the Cygnus constellation within 2.5$^{\circ}$ at 68 \% CL, corresponding to a nonambiguous detection of particles from
the Galactic halo which is in favor of a dark matter positive detection.\\
We found the same result as in \cite{billard.disco}, which was expected as the information on the main incoming direction 
is enclosed mainly in the angular part of the WIMP-induced spectrum. The major update is the fact that we prove that this result is
 model-independent as no {\it a priori} knowledge of neither the Galactic dark matter halo nor the WIMP particle 
is needed. The 2D correlations of Fig.~\ref{fig:fat50iso} (third and fourth columns and lines) and the two first  columns of the correlation matrix (see
Fig.~\ref{fig:corMatrix})
 indicate that there is no correlation of
the main incoming direction with the six other parameters.\\
 In addition, this result holds true for all cases studied hereafter where we have considered different input values of the 
 WIMP mass or halo model. Indeed, for each case,
 we have checked that the main incoming direction is always consistently constrained and reveals no correlation with other parameters (see Sec.~\ref{sec:param}).\\
We emphasize conclusions from \cite{billard.disco,green.disco} : directional detection of 
dark matter is a powerful strategy to clearly identify a positive dark matter signal, using the main incoming direction as  the discovery proof, 
even in the case of a sizeable background contamination and non standard halo model.

\subsection{Dark matter halo properties}
\label{sec:velocitydisp}
The originality of this work, in comparison to current phenomenological studies, is that the properties of the dark matter halo itself are constrained using 
a single directional detection experiment. 
As shown on Fig.~\ref{fig:fat50iso}, the velocity dispersions are  strongly and consistently constrained according to the input values. Indeed,
from the marginalized distributions of the posterior PDF of each velocity dispersion, the following constraints can be deduced 
\begin{eqnarray}
\sigma_x  & = & 158^{+15}_{-17} \ {\rm km.s^{-1}} \ (68\% \ {\rm CL}),  \nonumber \\
\sigma_y    & = & 164^{+27}_{-26} \ {\rm km.s^{-1}} \ (68\% \ {\rm CL}),  \nonumber \\
\sigma_z    & = & 145^{+14}_{-17} \ {\rm km.s^{-1}} \ (68\% \ {\rm CL}), \nonumber 
\end{eqnarray}
giving, in such case, strong evidence in favor of an isotropic dark matter halo.\\

However, one could notice, from the full MCMC result (Fig.~\ref{fig:fat50iso}) and from the correlation matrix (Fig.~\ref{fig:corMatrix}),
that the three velocity dispersions are quite correlated to each other, to the WIMP properties ($m_{\chi}, \log_{10}(\sigma_n)$) and to the background rate. In the following,
we propose a short discussion to understand the fundamental origin of these different correlations. To begin with, the positive correlation between each of the three 
velocity dispersions mainly comes from the information on the angular distribution. Indeed, in order to reproduce the shape of the velocity distribution, which is isotropic
in this case, the three velocity dispersions have to be positively correlated to each other; in this case, we found $\rho[\sigma_i,\sigma_j] \approx 0.4$ for $i\neq j \in$ \{x,y,z\}. 
However, increasing the velocity dispersions leads to an increase in the number of expected WIMP events $\mu_s$ and to wider WIMP event angular distribution. The latter can be compensated by
decreasing the WIMP mass as it leads to tighter angular distribution (see \cite{billard.disco,billard.exclusion} for a detailed discussion)
 thus implying a negative correlation between the WIMP mass and the three velocity dispersions with 
$\rho[m_{\chi},\sigma_x] = \rho[m_{\chi},\sigma_z] \approx -0.55$  and $\rho[m_{\chi},\sigma_y] \approx -0.75$. As the cross section is directly proportional to
 $\mu_s$ the correlations between the $\log_{10}(\sigma_n)$ and the three velocity dispersions are obviously negative with $\rho[\log_{10}(\sigma_n),\sigma_x] =
  \rho[\log_{10}(\sigma_n),\sigma_z] \approx -0.57$ and $\rho[\log_{10}(\sigma_n),\sigma_y] \approx -0.70$. Correlations between the WIMP parameters and $\sigma_y$ are stronger
  than in the case of $\sigma_x$ and $\sigma_z$ as it is the most related to $\mu_s$ and the total width of the angular event distribution. Then, as the velocity dispersion
  along the $y$ axis is more degenerated with the other parameters than $\sigma_x$ and $\sigma_z$ the error bar on the estimation of $\sigma_y$ are larger than for the two
  other velocity dispersions (about 2 times larger). The negative correlation between the three velocity dispersions and the background rate can be easily explained by the
  definition of the extended likelihood function where the sum of $\mu_s$ (proportional to the velocity dispersions) with $\mu_b$ (proportional to the background rate) 
  follows a Poisson distribution of mean equal to $\mu_s + \mu_b = N_{\text{event}}$; we found in this case: $\rho[R_{b},\sigma_j] \approx -0.4$ with $j \in$ \{x,y,z\}. \\

\begin{figure}[t]
\begin{center}
\includegraphics[scale=0.45,angle=0]{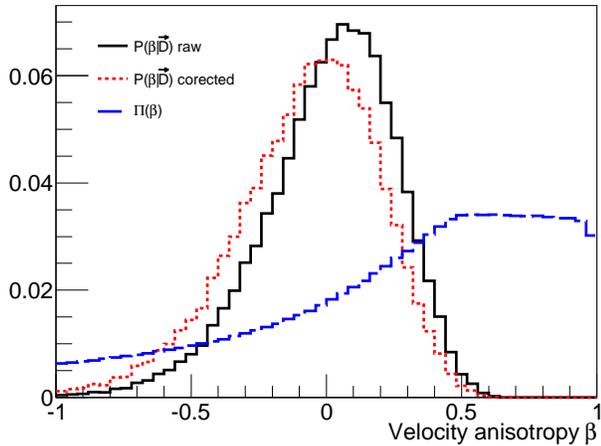}
\caption{Posterior PDF distribution of the $\beta$ parameter, with and without correction due to nonflat prior. The prior is Monte Carlo estimated.} 
\label{fig:beta}
\end{center}
\end{figure}

The evaluation of the velocity anisotropy parameter $\beta$ allows us to summarize the results from the three  velocity
dispersions.  Indeed, the posterior  PDF of the $\beta$ parameter can be computed from  Eq.\ref{eq:beta}.
However, having a flat prior on the three velocity dispersions implies a nonflat (informative) 
prior on the $\beta$ parameter. Hence, Fig.~\ref{fig:beta} presents the {\it raw} PDF of $\beta$ considering flat priors on the $\sigma_i$'s in the black solid line, 
the induced prior on the $\beta$ parameter $\Pi(\beta)$ is
shown as the blue dashed line, and the red dotted line corresponds to the {\it corrected} PDF of $\beta$  with a flat prior.  
In the following, for each case,
we will only consider the {\it corrected} posterior PDF of the $\beta$ parameter. From the latter, we can deduce an 
interesting constraint 
$\beta=-0.073^{+0.29}_{-0.18} \ (68\% \ {\rm CL})$ favoring an isotropic dark matter halo. 
This is a proof that  within 
the framework of the multivariate Gaussian halo model, a dedicated MCMC analysis of directional data would allow us to constrain 
the velocity dispersions,  resulting in a discrimination between various halo models.

\subsection{WIMP parameters}
\label{sec:WIMPParam}
As stated above, this MCMC analysis also allows us to constrain the parameters of the WIMP by considering both the angular and the energy information from each recoiling
event. 
Figure \ref{fig:fat50iso} (first 2 columns) presents marginalized distributions  and 2D correlation plots concerning the 
WIMP parameters ($m_{\chi}, \log_{10}(\sigma_n)$).    First, we can notice that this analysis method allows us to get satisfactory results, {\it i.e.} 
constraints which are consistent with the input values and with a rather small dispersion: 
\begin{align*}
& m_{\chi}   =  51.8^{+5.6}_{-19.4} \ {\rm GeV/c^2} \ (68\% \ {\rm CL}),  \\
& \log_{10}(\sigma_n)  =   -3.01^{+0.05}_{-0.08} \ (68\% \ {\rm CL})    
\end{align*}
Moreover, as the velocity dispersions are set as free parameters, 
 induced bias due to wrong halo model assumptions is avoided as long as the input halo model is consistent with our ansatz. 
 We refer the reader to \cite{green.masse1} 
 for a detailed discussion about the effect of halo model uncertainties on allowed regions. In fact, the combined use 
 of angular and energy  information allows us to remove degeneracies amongst the eight   parameters and hence to obviate bias in the determination of the WIMP properties.\\
We observe the usual strong correlation $\rho[m_{\chi},\log_{10}(\sigma_n)] \approx 1$ (see Fig.~\ref{fig:corMatrix}) between $m_{\chi}$ and $\log_{10}(\sigma_n)$ which is 
 inherent in the very definition of the event rate, as it scales basically with $\sigma_n/m_{\chi}$ for 
low mass target. We also found a small and positive correlation between the WIMP mass (inversely proportional to $\mu_s$) and the background rate (proportional to $\mu_b$)
 such as $\rho[m_{\chi},R_b] \approx 0.25$. Indeed, as mentioned
before, this correlation is straightforwardly due to the relationship between $\mu_s$ and $\mu_b$ where the total number of recorded events follows a Poisson distribution of mean 
 $\mu_s + \mu_b = N_{\text{event}}$. Finally, we found no correlation between $\log_{10}(\sigma_n)$ and the background rate $R_b$.\\
 
As a conclusion, directional detection provides a unique opportunity 
to constrain, with a single experiment,   the WIMP mass and the WIMP-nucleon cross section  within the framework of a high-dimensional multivariate analysis.
 This is of great interest in the context of phenomenological efforts 
 \cite{bernal2,green.masse1,green.masse2,Drees:2007hr,drees.masse,Chou:2010qt,Shan:2010hr,Shan:2010qv,Shan:2009ym,Bertone:2010rv} trying 
 to  constrain the WIMP parameters ($m_{\chi}, \sigma_n$) with upcoming dark matter experiments, with either an indirect, direct, or directional
strategy.  In this work, we have gone one step further in constraining
 the local WIMP velocity distribution with a single directional detection experiment.\\
  It is, of course, possible to include external data as nuisance parameters,
 {\it e.g.} measurement of the local dark matter density $\rho_0$ \cite{pdg,salucci}, the local circular velocity $v_0$ and the escape velocity (taken as infinity in this
 study). However, it seems premature at the level of a methodological study aimed at showing how to handle directional detection data. 
 For instance, the local WIMP  density  is usually quoted within the range  $\rho_0 \sim 0.2 - 0.8$ GeV.c$^{-2}$.cm$^{-3}$ and we used  
 the so-called  "standard" value
$\rm 0.3 \  GeV.c^{-2}.cm^{-3}$  for the sake of comparison with    various direct 
detector results  \cite{pdg}. We note that recently, a value of the local dark matter density, $\rho_0 = 0.43  \pm 0.11 \pm 0.10 \ {\rm   GeV.c^{-2}.cm^{-3}}$, has 
been derived within the framework of a Galaxy-model-independent
method \cite{salucci}.  All constraints on  the WIMP-nucleon cross section  can be relaxed  into  constraints on $\rho_0 \times \sigma$.

\subsection{Background estimation}
The background rate estimation is also a key point of this analysis strategy, not for the value itself but for the fact that  a wrong
background estimation may induce bias for other parameters.  Indeed, as upcoming data will necessarily be contaminated by some background events,
 it is important to be able to manage them. 
As shown in Fig.~\ref{fig:fat50iso} (last row), it is correctly estimated from the MCMC : 
$R_b = 10.97 \pm 1.2 \ {\rm kg^{-1}year^{-1}} \ (68\% \ {\rm CL})$, with tiny
correlations with other parameters already discussed in the previous sections.  Then, the fact that the 
background rate is left as a free parameter and reconstructed with the MCMC method allows us to avoid bias in the estimation of the 
other parameters.
Qualitatively, the background rate is mainly constrained by the angular part of the spectrum, more precisely in the hemisphere opposite to
the Cygnus constellation, where few WIMP events are expected. In fact, the quality of the estimation of the 
WIMP and halo parameter is directly related to the estimation of the background rate. 
In this example,  we have shown that dark matter parameter estimation (main direction, WIMP, and dark matter halo properties) is not affected by a rather large 
background fraction ($\sim 30\%$). Hence, directional detection can accommodate to a sizeable background contamination (posterior to data selection),  suggesting the idea that light shielding
 might be sufficient, thus allowing us to reduce  
muon-induced neutron background \cite{Mei:2005gm}.\\
As stated above, for this example a flat background energy spectrum has been considered, which is indeed an optimistic case. 
In Sec.~\ref{sec:param}, we study the effect of considering an energy distributions for  background events which is
 similar to the one for WIMP events.

\section{Results for various input models}
\label{sec:param}

The constraints on the different parameters obviously depend on the input model, characterized by
 the WIMP and dark matter halo properties as well as the 
background energy spectrum. Indeed, the directional WIMP event rate crucially depends on the dark matter parameters, both from particle physics and
Galactic halo physics, and degeneracies may arise
depending on their input values. In the following, we explore various input models in order to evaluate their impact on the different constraints
which could be obtained with a single directional detection experiment, as the one proposed by the MIMAC collaboration, using our MCMC analysis. \\

 \begin{figure}[t]
\begin{center}
\includegraphics[scale=0.42,angle=0]{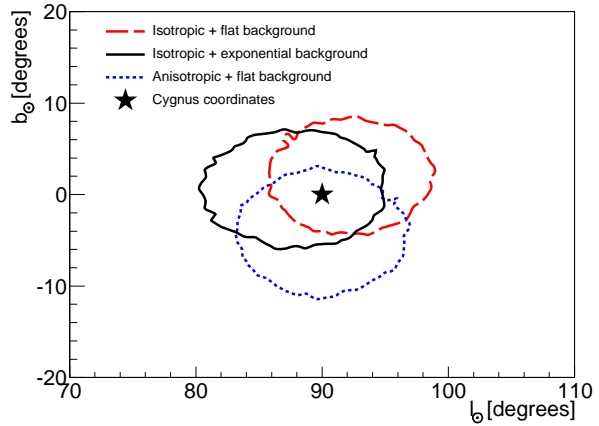}
\caption{95\% contour level in the ($\ell_{\odot},b_{\odot}$) plan  for three input models: Isotropic halo model + exponential background (solid line), Isotropic halo model + flat
background (long dashed line) and anisotropic halo model + flat background (dotted line).} 
\label{fig:discovariousmodel}
 \end{center}
\end{figure}

The first point worth emphasizing is the fact that in all cases presented hereafter, the recovered main recoil 
direction is always pointing towards Cygnus, within at most $\sim 4 ^{\circ}$ at 95\% CL (see Fig.~\ref{fig:discovariousmodel}).
 This is relatively straightforward, given the fact 
that this directional signature  is uncorrelated with the other parameters of
the MCMC analysis, as emphasized in Sec.~\ref{sec:proof}. Indeed, it has been shown in \cite{billard.disco} that this directional signature only
 depends on 
the background contamination, which is taken equal to 10 evts/kg/year in every following cases.
 This outlines the robustness of the choice of this parameter as a relevant observable
to prove that a positive detection of dark matter has been reached by a directional detector.  As outlined in  
\cite{billard.disco}, this would allow directional detection to provide evidence in favor
of a detection of Galactic dark matter even at low exposure and even with a sizeable background contamination. In this study, we have  checked that this conclusion holds 
true even in the case of non standard dark matter halo model.

\begin{figure*}[p]
\begin{center}

\includegraphics[scale=0.4,angle=0]{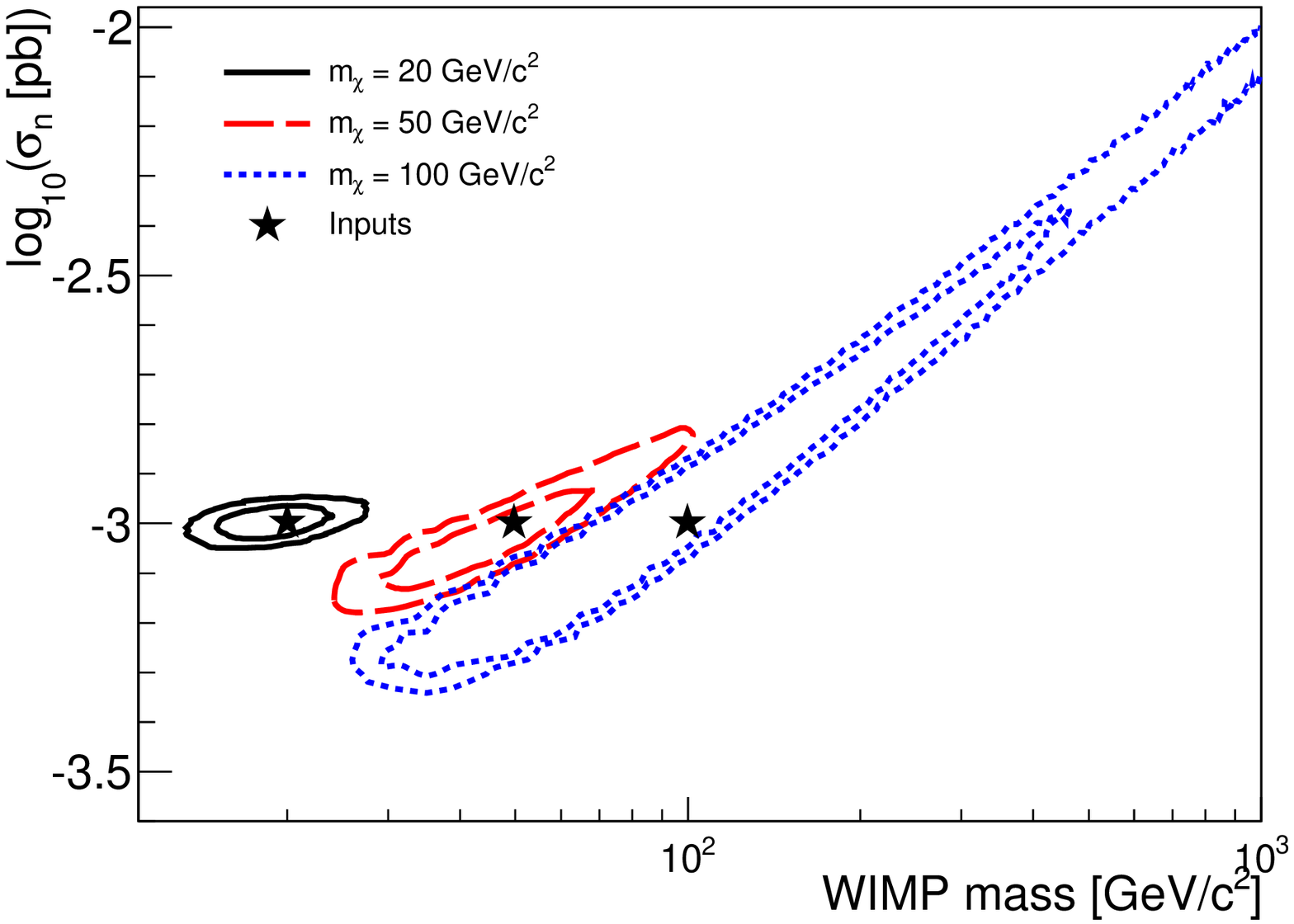}
\includegraphics[scale=0.4,angle=0]{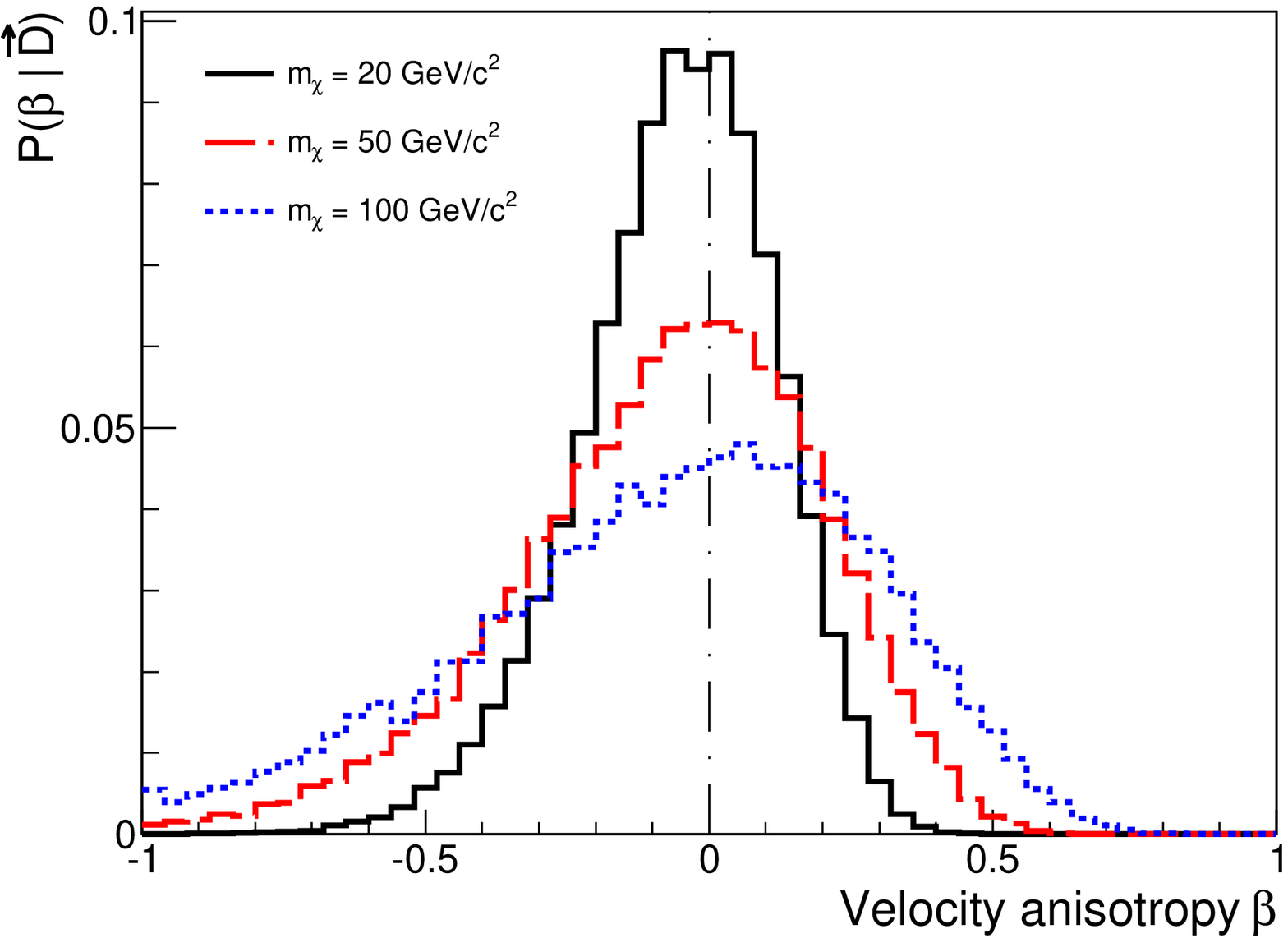}
\caption{Left panel : 68\% and 95\% contour level in the ($m_{\chi},\sigma_n$) plane, for the isotropic input model and for a WIMP mass 
equal to 20, 50 and 100 $\rm GeV/c^2$. 
Right panel : posterior PDF distribution of the $\beta$ parameter for the same models.} 
\label{fig:WIMPMass}

\includegraphics[scale=0.4,angle=0]{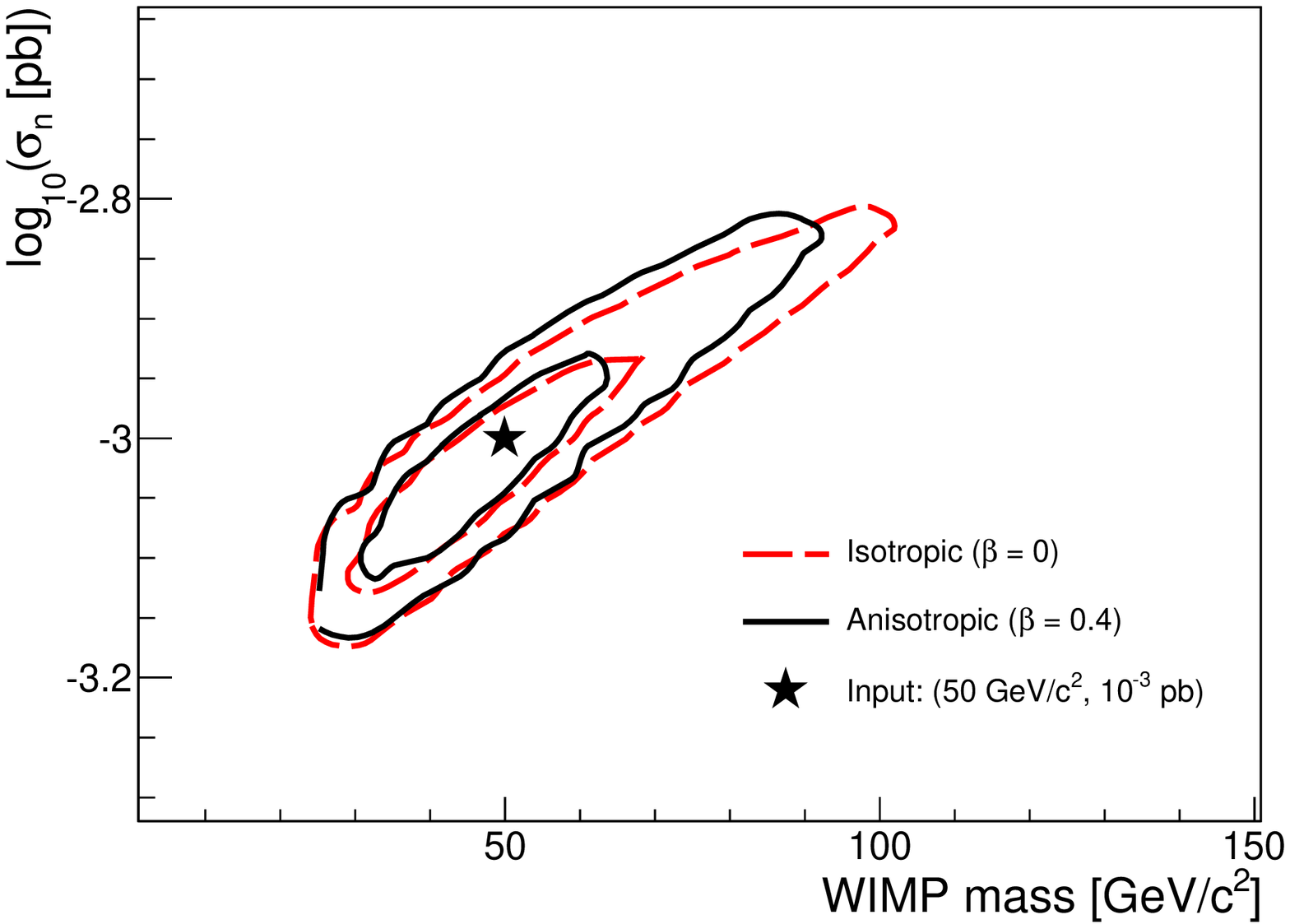}
\includegraphics[scale=0.4,angle=0]{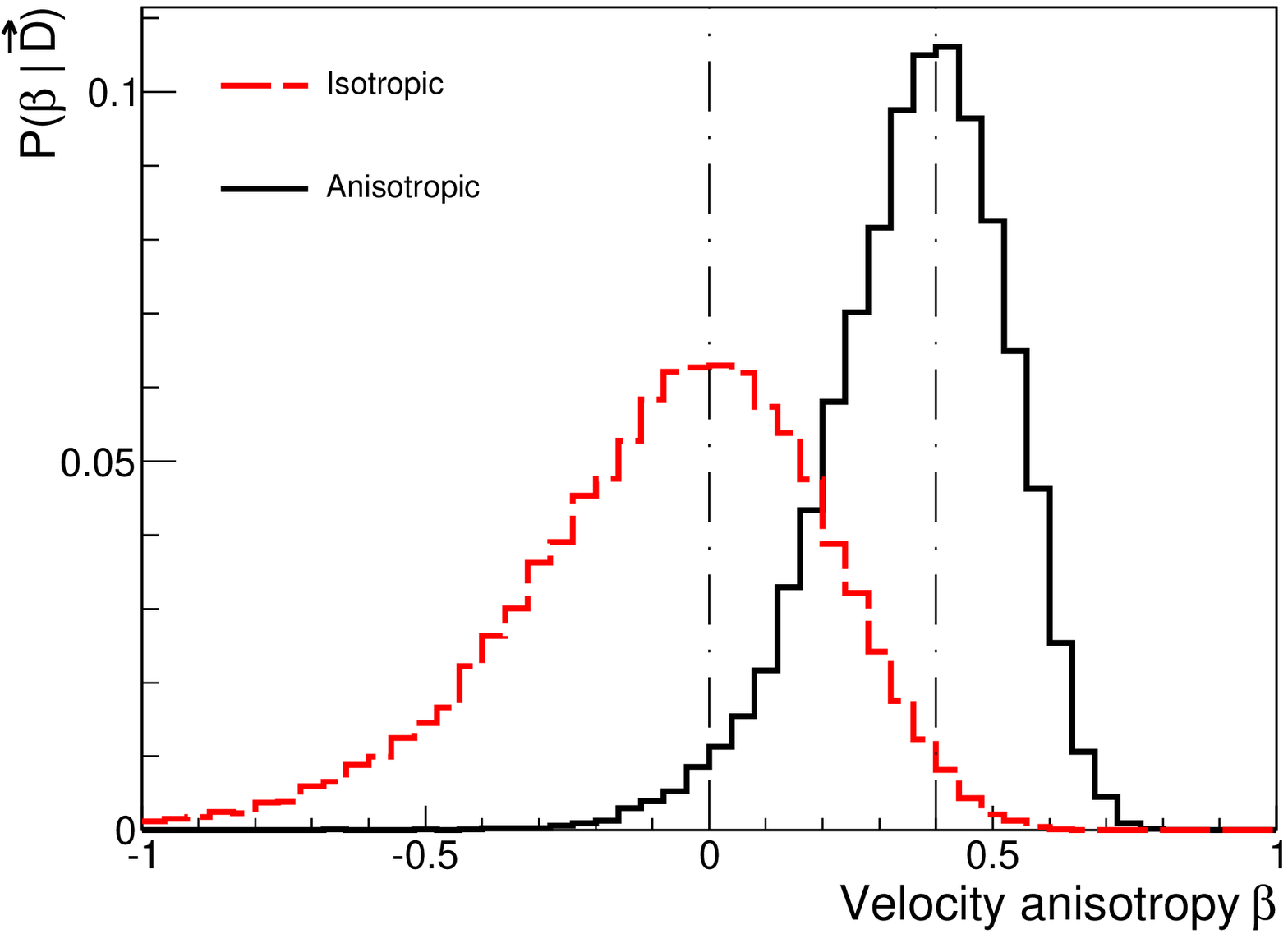}
\caption{Left panel : 68\% and 95\% contour level in the ($m_{\chi},\sigma_n$) plane, for a 50 $\rm GeV/c^2$ WIMP and for two input models : isotropic ($\beta=0$) and 
triaxial ($\beta=0.4$). 
Right panel : posterior PDF distribution of the $\beta$ parameter for the same models.}  
\label{fig:HaloT4}

\includegraphics[scale=0.4,angle=0]{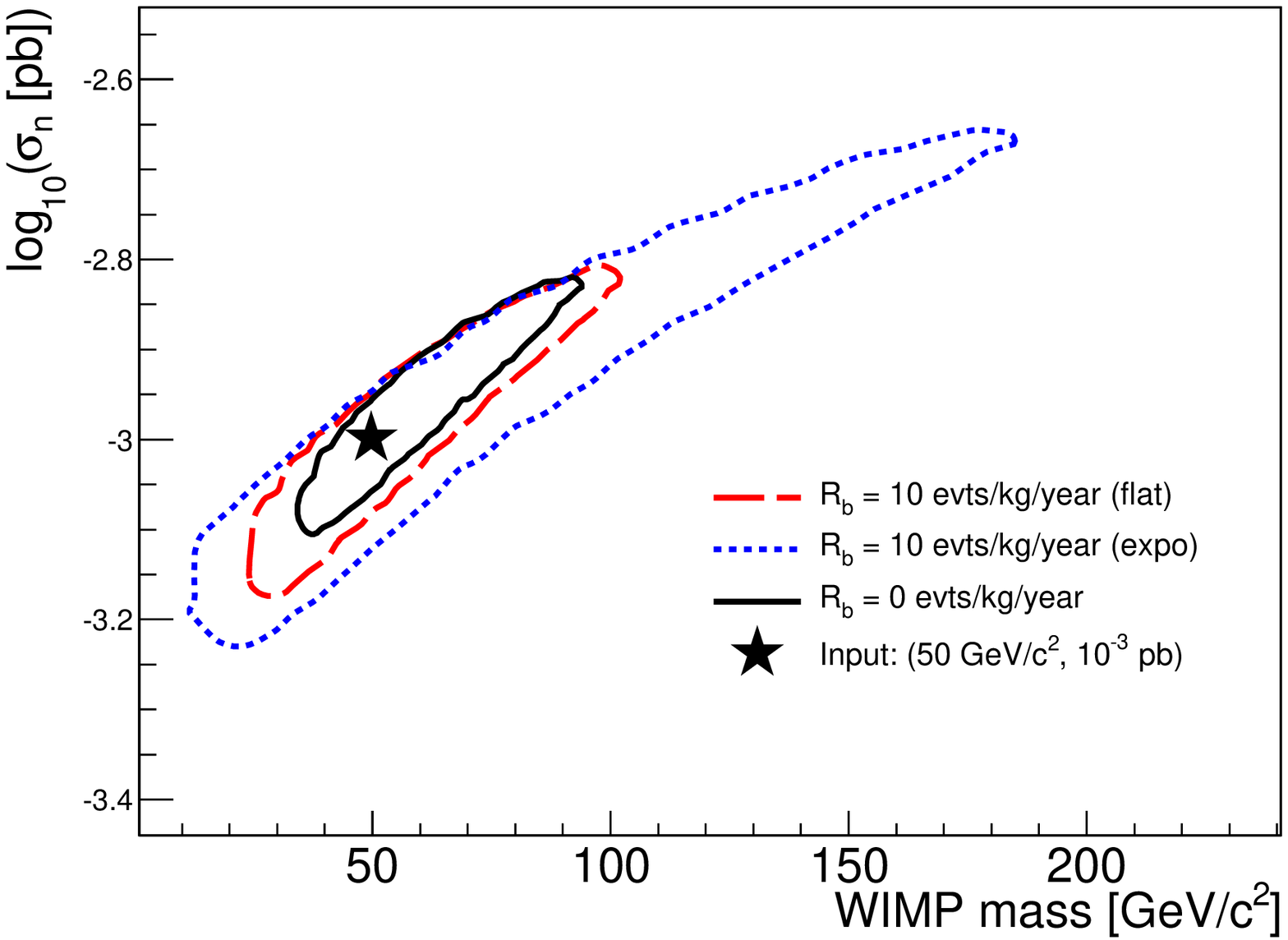}
\includegraphics[scale=0.4,angle=0]{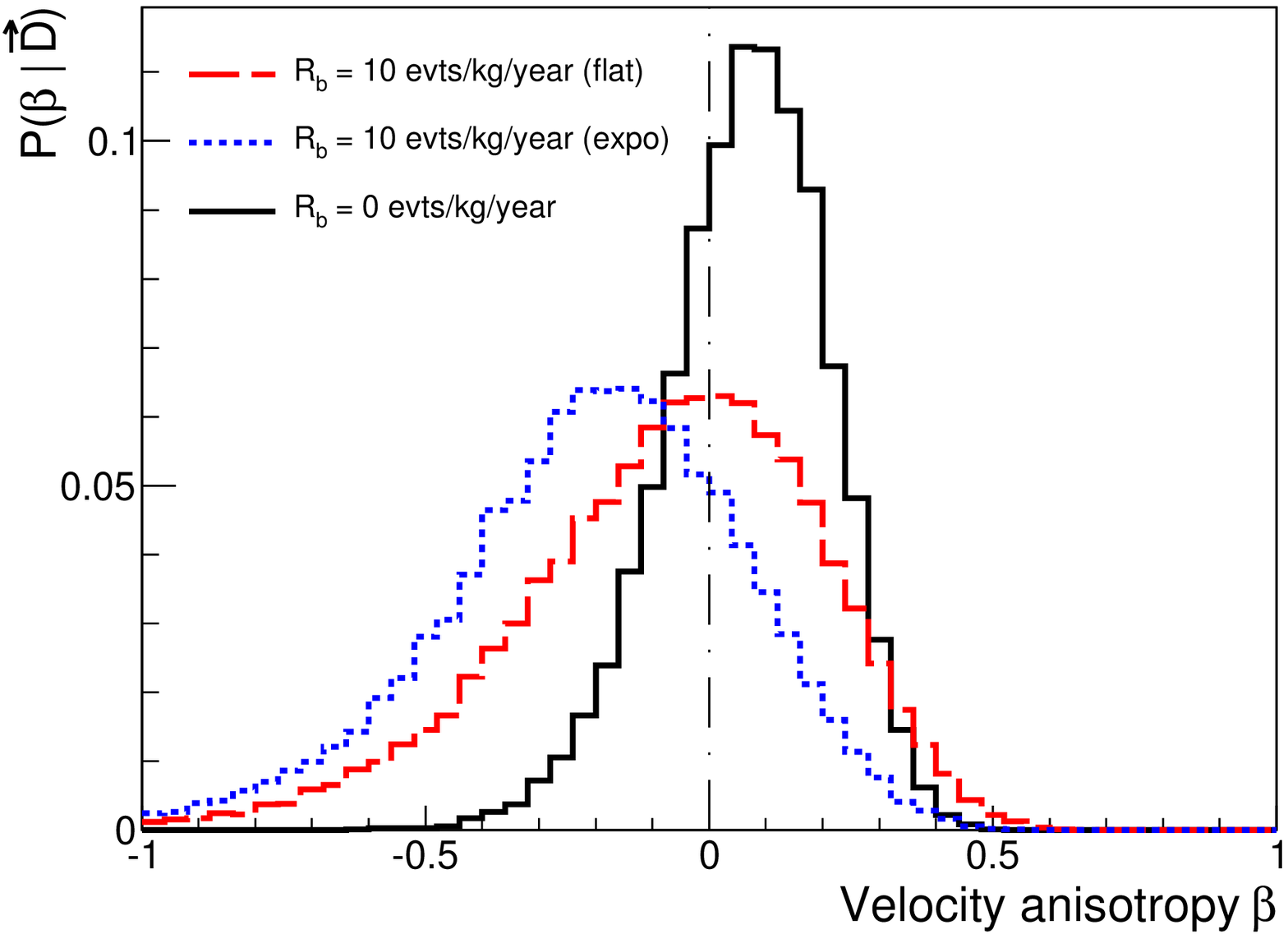}
\caption{Left panel : 95\% contour level in the ($m_{\chi},\sigma_n$) plane, for a 50 $\rm GeV/c^2$ WIMP and for three input background model : no background, flat spectrum and  
exponential spectrum. 
Right panel : posterior PDF distribution of the $\beta$ parameter for the same models.} 
\label{fig:Background}
\end{center}
\end{figure*}

\subsection{Varying the input WIMP mass}
As highlighted by several previous studies \cite{green.masse1,green.masse2,bernal2,Bertone:2010rv}, the WIMP mass plays a key role in 
the shape of the allowed regions. We have simulated
three different sets of directional data corresponding to an input WIMP mass of 
$m_{\chi} = 20, 50, 100 \ {\rm GeV/c^2}$ with a constant WIMP-nucleon cross section  $\sigma_n = 10^{-3} \ {\rm pb}$,
considering a MIMAC-like directional detector (Sec.~\ref{sec:directional}) and the standard isotropic halo model. 
The results from the three MCMC runs are illustrated in Fig.~\ref{fig:WIMPMass}. 
We present for the three WIMP masses, in the left panel,  the 68\% and 95\% CL contours in the 
($m_{\chi},\log_{10}(\sigma_n)$) plane, and in the right panel, the posterior PDF $P(\beta|\vec{D})$ of 
the anisotropy velocity parameter $\beta$.\\
The WIMP properties ($m_{\chi}, \log_{10}(\sigma_n)$) are consistently constrained according to the input values with no {\it a priori} knowledge of 
the halo properties, as the velocity dispersions are set as free parameters of the analysis. It can be deduced from Fig.~\ref{fig:WIMPMass} 
that  this analysis is working for any input WIMP mass even if the constraints strongly depend on the
input value. Indeed, as it can be seen in Fig.~\ref{fig:WIMPMass}, the constraints on ($m_{\chi},\log_{10}(\sigma_n)$) are very tight below 50 GeV/$c^2$ and become wider 
 for increasing WIMP mass.\\
In fact, for the
 100 GeV/c$^2$ input WIMP mass, only a lower limit should be deduced as $m_{\chi} > 30$ GeV/c$^2$ (68\% C.L.). Indeed, 
 the 68\% and 95\% C.L. contours correspond to the case where a flat prior on $m_{\chi} \in [5,10^3]$ GeV/c$^2$ is considered.
These weaker constraints in the case of a heavy WIMP are due to the fact that the signal characteristics, {\it i.e} the slope of  the 
energy distribution and the width of the angular distribution, evolve slowly with the WIMP mass once $m_{\chi} \geq 100$ GeV/c$^2$ for 
a fluorine target and a
 recoil energy in the range  [5,50] keV, as shown in \cite{billard.disco}.\\
As a consequence of this weaker constraint at heavy WIMP masses, the constraints on the halo properties are also getting 
weaker, albeit with smaller effect. Indeed, as  shown in the right panel
of Fig.~\ref{fig:WIMPMass}, the constraint on the anisotropy velocity parameter $\beta$ is stronger (smaller error bars) 
for an input WIMP mass of 20 GeV/c$^2$  than for  
a 100 GeV/c$^2$ one. However, as highlighted in Table~\ref{tab:inout}, the constraint on the $\beta$ parameter remains competitive and 
for a  30 kg.year exposure with a MIMAC like directional detector, this MCMC would allow us to get, in this case,  
a strong evidence in favor of an isotropic dark matter halo.\\

\setlength{\tabcolsep}{0.1cm}
\renewcommand{\arraystretch}{1.4}
\begin{table}[t]
\begin{center}
\begin{tabular}{|c|c|c|c|c|}
\hline
 Halo & Background &  $m_{\chi} \ {\rm (GeV/c^2)}$ &   $\beta_{in}$ & $\beta_{out}$   \\ \hline \hline
Isotropic & Flat   &  20  & 0  &    $-0.06^{+0.2}_{-0.1}$ \\ \hline
Isotropic & Flat   &  50  & 0  &    $-0.07^{+0.3}_{-0.2}$ \\ \hline
Isotropic & Expo.  &  50  & 0  &    $-0.20^{+0.3}_{-0.2}$ \\ \hline
Isotropic & NO     &  50  & 0  &    $+0.05^{+0.1}_{-0.1}$ \\ \hline 
Isotropic & Flat   &  100 & 0  &    $-0.10^{+0.4}_{-0.2}$ \\ \hline \hline
Anisotropic & Flat &  50  & 0.4  &  $+0.38^{+0.2}_{-0.1}$ \\ \hline  
\end{tabular}
\caption{Values of the $\beta$ parameter from the marginalized distribution 
for various input models. 
We quote mean value of the PDF distribution and 68 \% CL error bars.}
\label{tab:inout}
\end{center}
\end{table}
\renewcommand{\arraystretch}{1.1}

\subsection{Effect of an anisotropic input halo model}
In this section, we   vary the input halo model to evaluate the evolution of the constraints associated with the different 
dark matter properties  ($m_{\chi},\sigma_n,\beta$). Indeed,  as the velocity dispersions are set as free parameters, 
induced bias due to wrong model assumption should be avoided. This is for instance the effect observed in  \cite{green.masse1}, 
with a systematic downward shift of the estimated cross section, when assuming a    standard isotropic 
 velocity distribution fitting model whereas the input model is a triaxial one.\\
In the following, we investigate the effect of an extremely triaxial input halo model with $\beta=0.4$
 (see sec.\ref{halomodeling}) on the estimation   of the dark matter parameters.\\

The results from the MCMC run on a simulated 
dataset corresponding to a WIMP mass of 50 GeV/c$^2$ with the latter anisotropic halo model are presented in
Fig.~\ref{fig:HaloT4}. As for the previous section, in the left panel is presented the constraint at 68\% and 95\% on the 
($m_{\chi},\log_{10}(\sigma_n)$) plane, while
 in the right panel is given the
deduced posterior PDF of the $\beta$ parameter. For convenience and comparison, 
the results from the benchmark input model (isothermal sphere with a 50 GeV/c$^2$ WIMP) are recalled.\\
From the left panel of Fig.~\ref{fig:HaloT4}, we can conclude that the two halo models give similar constraints which are both consistent with the input values.
 In fact, and as foreseen, the fact that the velocity
dispersions are set as free parameters in the MCMC analysis allows us to avoid induced bias due to wrong model assumption.\\

From the right panel of Fig.~\ref{fig:HaloT4} we can deduce that the $\beta$ parameter is well constrained: 
$\beta = 0.38^{+0.2}_{-0.1}$, as in the isotropic case. In fact,  the constraint is even stronger in the anisotropic case
 than in the isotropic one. This comes straightforwardly from the decrease of the degeneracy  between the three velocity dispersions with increasing
 departure from isotropy.\\
As a conclusion of this study, it should be highlighted that the combination of   information from 
the angular and energy distributions leads to robust allowed regions in the ($m_{\chi},\log_{10}(\sigma_n)$) plane, since 
the halo model is also being constrained with the MCMC analysis from the same dataset of a single directional detection experiment. 
 Moreover, the velocity anisotropy parameter $\beta$, {\it i.e.} the three velocity dispersions,
  could be sufficiently constrained to discriminate between different halo  models with future directional detectors such as the one proposed by the MIMAC collaboration
  \cite{mimac}.

\subsection{Varying the input background spectrum}
The background energy spectrum is a key issue for both directional detection and direct detection 
(direction-insensitive experiments). When setting exclusion limits with directional detection, 
the difficulty can be avoided by considering only the angular part of the directional event rate, 
thus allowing us to set  robust and conservative limits \cite{billard.exclusion}. But as far as  
the whole directional event rate is used, the question of the background energy spectrum 
must be carefully treated. In fact, as the background energy spectrum is unknown it must be guessed to be included in a likelihood-type
analysis. Then, a wrong assumption on the background shape leads to an incorrect estimation of the background rate, resulting in 
a wrong estimation of the dark matter properties.\\
Motivated by  simulations of neutron background in  dark matter detectors, {\it e.g.} in low pressure TPC \cite{Carson:2005qz},  
 two different background energy distributions are usually considered  \cite{bernal2,green.masse1,green.masse2,Drees:2007hr,drees.masse,Shan:2010hr} 
 : flat and exponentially decreasing with increasing recoil energy. The exponential one corresponds  to 
 the most pessimistic case as it is chosen, in our case, to be exactly the same as the WIMP-induced energy distribution. That is to say  an exponential distribution
  with a slope of
 $\sim$ 17 keV in the case where the WIMP mass is 50 GeV/c$^2$ and considering an isotropic halo model. 
 As outlined in \cite{green.masse2}, it is not possible to disentangle a 
WIMP signal from the background, with a single direct detector,  if the shape of the background and WIMP-induced energy distributions are similar. In principle,
this will not be the case for directional detection as the angular distribution of the background is isotropic, 
then remaining different from  the WIMP-induced one.\\

\setlength{\tabcolsep}{0.1cm}
\renewcommand{\arraystretch}{1.4}
\begin{table}[t]
\begin{center}
\begin{tabular}{|c|c|c|c|}
\hline
 Halo & Background &  $R_b^{in}$ &   $R_b$ MCMC output  \\  \hline \hline
Isotropic   & Flat     & 10  &    $10.97^{+1}_{-1}$ \\ \hline
Isotropic   & Expo.    & 10  &    $10.03^{+3}_{-2}$ \\ \hline
Isotropic   & NO       &  0  &    $< 0.36$ (upper limit) \\ \hline   \hline 
Anisotropic & Flat     &  10  &   $9.8^{+1}_{-1}$ \\ \hline  
\end{tabular}
\caption{Values of the $R_b$ parameter (in $\rm kg^{-1}year^{-1}$) from the marginalized distribution 
for various background input models. 
We quote mean value of the PDF distribution while the error bars and upper limits are quoted with a 68 \% CL.}
\label{tab:inout-b}
\end{center}
\end{table}
\renewcommand{\arraystretch}{1.1}

Figure \ref{fig:Background} presents the constraints in the ($m_{\chi},\log_{10}(\sigma_n)$) plane and on the $\beta$ parameter for three input 
background energy distributions: no background (black solid line), flat (red dashed line) and exponential (blue dotted line).
 The other input parameters are those used in Sec.~\ref{sec:res.iso} for the benchmark input model (an isotropic
halo, a $50 \ {\rm GeV/c^2}$ WIMP mass and a $10^{-3} \ {\rm pb}$ WIMP-nucleon axial cross section).\\
First, the comparison between the case with no background and the case with a flat background energy distribution highlights the fact that even with a large background
contamination ($\sim 30\%$), the results are quite similar, particularly in the determination of the WIMP properties,
 due to the fact that the disentanglement between WIMP and background events is done with both energy and directional arguments.
Then, in both cases, as the background rate is correctly estimated by the MCMC analysis 
(see Table~\ref{tab:inout-b}), systematic bias in the estimation of the dark matter properties is avoided. However, from Fig.~\ref{fig:Background} and
Table~\ref{tab:inout}, it should be noticed that, even if the $\beta$ parameter is consistently constrained according to the input value, the presence of a sizeable background
leads to a wider constraint (about two times larger).\\
 
In the case of an exponential input background energy distribution,   the result is basically unchanged, although 
constraints are  weaker.  This is due to the fact that the background 
event rate parameter is less constrained (see   Table~\ref{tab:inout-b}) 
resulting in broader   marginalized distributions of other parameters. 
Indeed, the estimation of the $R_b$ parameter is done solely with the angular part of the spectrum, as 
 the energy distributions are exactly the same for both kinds of events. 
  Nevertheless, even in such a pessimistic case, the WIMP properties 
 ($m_{\chi},\log_{10}(\sigma_n)$) and the dark matter halo properties encoded in the $\beta$ parameter can still be estimated
  with upcoming directional detectors with realistic exposures thanks to the use of this MCMC analysis.\\

From this study, it should be concluded that, the effect of background contamination on directional data can be handled in the case where the background energy distribution
is correctly estimated. 
Eventually, we have shown that even for a large background contamination and in  the most pessimistic background model, directional detection 
combined with this MCMC analysis should allow us  to assess consistent and interesting constraints on the dark matter properties with a single experiment.


\section{Conclusion}
We have shown that 
identification of dark matter might be achieved with a  10 kg $\rm CF_4$ directional detector, allowing  3D track reconstruction with sense recognition down to 5 keV and  
 operated during three years. To fully exploit upcoming data, we propose a new high dimensional multivariate analysis
method  based on  a Markov chain Monte Carlo analysis of recoil events, allowing  to 
constrain, in a single directional experiment, the  WIMP parameters, both from particle physics (mass and cross section) 
and Galactic halo (velocity dispersion along the three axis) and within the framework of a given ansatz.

Indeed, the combination of   information from 
the angular and energy distributions leads to robust allowed regions in the ($m_{\chi},\log_{10}(\sigma_n)$) plane, since 
the halo model is also being constrained with the MCMC analysis from the same dataset of a single directional detection experiment. 
 Moreover, the velocity anisotropy parameter $\beta$, related to the three velocity dispersions,
  could be sufficiently constrained to discriminate between various halo  models with future directional detectors such as the one proposed by the MIMAC collaboration
  \cite{mimac}.

\section*{Acknowledgements}
The authors would like to thank Antje Putze and Laurent Derome for fruitful discussions concerning MCMC methods.

\end{document}